\documentclass[]{aastex631}

\usepackage{rotating}

\received{\today}

\submitjournal{PSJ}

\shorttitle{45P/HMP and 46P/Wirtanen Dust Radial Profiles}
\shortauthors{Lejoly et al.}

\graphicspath{{./}{figures/}}

\begin{document}

\title{Radial Distribution of the Dust Comae of Comets 45P/Honda-Mrkos-Pajdu\u s\'akov\'a and 46P/Wirtanen}

\author{C. Lejoly}
\affiliation{Lunar and Planetary Laboratory \\
1629 E University Blvd \\
Tucson AZ 85721-0092, USA }

\author{W. Harris}
\affiliation{Lunar and Planetary Laboratory \\
1629 E University Blvd \\
Tucson AZ 85721-0092, USA }

\author{N. Samarasinha}
\affiliation{Planetary Science Institute \\
1700 East Fort Lowell, Suite 106 \\
Tucson, AZ 85719-2395, USA }

\author{B.E.A. Mueller}
\affiliation{Planetary Science Institute \\
1700 East Fort Lowell, Suite 106 \\
Tucson, AZ 85719-2395, USA }

\author{E. Howell}
\affiliation{Lunar and Planetary Laboratory \\
1629 E University Blvd \\
Tucson AZ 85721-0092, USA }

\author{J. Bodnarik}
\affiliation{Lunar and Planetary Laboratory \\
1629 E University Blvd \\
Tucson AZ 85721-0092, USA }

\author{A. Springmann}
\affiliation{Lunar and Planetary Laboratory \\
1629 E University Blvd \\
Tucson AZ 85721-0092, USA }

\author{T. Kareta}
\affiliation{Lunar and Planetary Laboratory \\
1629 E University Blvd \\
Tucson AZ 85721-0092, USA }

\author{B. Sharkey}
\affiliation{Lunar and Planetary Laboratory \\
1629 E University Blvd \\
Tucson AZ 85721-0092, USA }

\author{J. Noonan}
\affiliation{Lunar and Planetary Laboratory \\
1629 E University Blvd \\
Tucson AZ 85721-0092, USA }

\collaboration{14}{(4*P Campaign)}
\author{L. R. Bedin}
\affiliation{INAF - Osservatorio Astronomica di Padova \\
Vicolo dell'Osservatorio 5 \\
35122 Padova,  Italy}

\author{J.-G. Bosch}
\affiliation{SpaceObs \\
San Pedro de Atacama - Chile \\
https://www.spaceobs.com}

\author{A. Brosio}
\affiliation{L'osservatorio Astronomico di Savelli \\
Via Villaggio Pino Grande (SP 28) \\
Località Timpa Caccianinni, 88825, Savelli (KR),  Italy}

\author{E. Bryssinck}
\affiliation{CARA-consortium, (Cometary ARchive for Afrho) http://cara.uai.it/home}
\affiliation{Vereniging Voor Sterrenkunde (VVS), \\
Oostmeers 122 C, \\
8000 Brugge, Belgium ( https://www.vvs.be )}
\affiliation{BAA, The British Astronomical Association, \\
Burlington House, Piccadilly, \\
London, W1J 0DU, United Kingdom}

\author{J.-B. de Vanssay}
\affiliation{SpaceObs \\
San Pedro de Atacama - Chile \\
https://www.spaceobs.com}

\author{F.-J. Hambsch}
\affiliation{American Association of Variable Star Observers (AAVSO),\\
49 Bay State Road, \\
Cambridge, MA 02138, USA }
\affiliation{Vereniging Voor Sterrenkunde (VVS), \\
Oostmeers 122 C, \\
8000 Brugge, Belgium ( https://www.vvs.be )}

\author{O. Ivanova}
\affiliation{Astronomical Institute of the Slovak Academy of Sciences,\\
Slovakia}
\affiliation{Main Astronomical Observatory of the National Academy of Sciences of Ukraine,\\
Ukraine}
\affiliation{Taras Shevchenko National University of Kyiv, \\
Astronomical Observatory, Ukraine }

\author{V. Krushinsky}
\affiliation{ Laboratory of Astrochemical Research,\\
Ural Federal University,\\
Ekaterinburg, Russia, ul. Mira d. 19, \\
Yekaterinburg, Russia, 620002\\}

\author{Z.-Y. Lin}
\affiliation{National Central University \\
No. 300, Zhongda Rd. \\
Zhongli District, Taoyuan City 32001, Taiwan (R.O.C.)}

\author{F. Manzini}
\affiliation{Stazione Astronomica di Sozzago \\
28060 Sozzago (Novara), Italy}

\author{A. Maury}
\affiliation{SpaceObs \\
San Pedro de Atacama - Chile \\
https://www.spaceobs.com}

\author{N. Moriya}
\affiliation{siOnet Ltd. - Applied Modeling Research\\
Faran Observatory, \\
Mitzpe-Ramon, Negev 47113, Israel}

\author{P. Ochner}
\affiliation{Dept. of Physics and Astronomy \\
University of Padova \\
Via Marzolo 8, \\
35131 Padova,  Italy}
\affiliation{INAF - Osservatorio Astronomica di Padova \\
Vicolo dell'Osservatorio 5 \\
35122 Padova,  Italy}

\author{V. Oldani}
\affiliation{Stazione Astronomica di Sozzago \\
28060 Sozzago (Novara), Italy}

\begin{abstract}

There was an unprecedented opportunity to study the inner dust coma environment, where the dust and gas are not entirely decoupled, of comets 45P/Honda-Mrkos-Pajdu\u s\'akov\'a (45P/HMP) from Dec. 26, 2016 - Mar. 15, 2017, and 46P/Wirtanen from Nov. 10, 2018 - Feb. 13, 2019, both in visible wavelengths.  The radial profile slopes of these comets were measured in the R and HB-BC filters most representative of dust, and deviations from a radially expanding coma were identified as significant. The azimuthally averaged radial profile slope of comet 45P/HMP gradually changes from $-1.81 \pm 0.20$ at 5.24 days pre-perihelion to $-0.35 \pm 0.16$ at 74.41 days post perihelion. Contrastingly, the radial profile slope of 46P/Wirtanen stays fairly constant over the observed time period at $-1.05 \pm 0.05$. Additionally, we find that the radial profile of 46P/Wirtanen is azimuthally dependent on the skyplane-projected solar position angle, while that of 45P/HMP is not. These results suggest that comet 45P/HMP and 46P/Wirtanen have vastly different coma dust environments and that their dust properties are distinct. As evident from these two comets, well-resolved inner comae are 
vital for detailed characterization of dust environments.

\end{abstract}

\section{INTRODUCTION} \label{sec:intro}

From 2016-2019, there was a unique opportunity to study the dust and gas environment in the inner coma - typically defined as several 1000 kms from the nucleus - of three closely approaching comets, 41P/Tuttle-Giacobini-Kres\`ak, 45P/Honda-Mrkos-Pajdu\u s\'akov\'a  (45P/HMP), and 46P/Wirtanen. In this paper, we focus on the latter two comets. Both comets 45P/HMP and 46P/Wirtanen approached Earth to within 0.08 au, and were well-placed for observational study. The close approach of these Jupiter Family Comets (JFCs) allowed for high spatial resolutions of the inner comae, which is a region typically not well resolved, except by spacecraft or during the rare occasions when a comet has a close encounter with Earth. The proximity we are obtaining for these comets simply due to their orbit's close approach to Earth may not resolve them quite as well as a close flyby or in situ study, but does provide a much more in depth opportunity, without requiring the typical cost of a spacecraft mission. Table \ref{perigee_distances} shows the heliocentric and geocentric distance ranges, and the perihelion and perigee distances for our observing spans \citep{Horizon}.

Comet 45P/HMP is a JFC with a perihelion distance of 0.53 au and an orbital period of 5.25 years. It was found to have a radius of 600-650 km and a rotation period of $\sim$7.5 hours \citep{Lejoly}. Comet 46P/Wirtanen is also a JFC, with an orbital period of 5.4 years and a perihelion distance of 1 au. It had a close approach to Earth on 16 December 2018 at 0.08 au and was well placed for long term monitoring from ground based observation. With its small radius of 0.6 km \citep{Lamy}, it has been described as an hyperactive comet, meaning that its activity level is higher than its expected active fraction of the nucleus. \citet{Farnham} found that it had a period of around 9 hours during its 2018 passage.

\begin{table}[h]
\centering
\caption{Geometric parameters for comets 45P/HMP and 46P/Wirtanen. For the Heliocentric and geocentric distances for our observations, negative values mean pre-perihelion/perigee. The individual observations are provided in Tables \ref{45P_dates_table} \& \ref{46P_dates_table}. }
\label{perigee_distances}
\begin{tabular}{|l|c|c|c|c|c|c|c|}
\hline
\textbf{Comet}    & \multicolumn{1}{m{2cm}|}{\textbf{Perihelion Date \& Time (UT)}} & \multicolumn{1}{m{2cm}|}{\textbf{Perihelion Distance (au)}}  & \textbf{r\footnote{Heliocentric range} (au)} & \multicolumn{1}{m{2cm}|}{\textbf{Perigee Date (UT)}}& \multicolumn{1}{m{2cm}|}{\textbf{Perigee Distance (au)}} & \textbf{$\Delta$\footnote{Geocentric range} (au)}  \\ \hline
\textbf{45P/HMP}         & 2016-12-31 06:29 & 0.53  &  -0.54 $\to$ 1.43  & \textbf{2017-02-11} 7:03   &  0.08       & -0.82 $\to$ 0.47  \\ \hline

\textbf{46P/Wirtanen}    & 2018-12-12 22:20 & 1.06   &  -1.14 $\to$ 1.34  & 2018-12-16 2:10 &  0.08         & -0.22 $\to$ 0.41                \\ \hline
\end{tabular}
\end{table}

\subsection{Cometary Dust \label{importance_dust}}

Cometary dust is defined as an ``unequilibrated, heterogeneous mixture of minerals, including both high- and low-temperature condensates" \citep{CometsII_dust} that does not typically sublimate upon close approach to the sun, but that rather can be dragged along with the sublimating gases. Cometary dust is a primordial part of a comet, being mostly unaltered particles containing both pre-solar grains and solar nebula condensates. Additionally, the Stardust mission, which brought samples back from the coma of comet 81P/Wild 2, detected Calcium Aluminum-rich Inclusions, suggesting some cometary dust was created in the hot protoplanetary disk \citep{Brownlee_2014}. Cometary dust measurements show a mixture of rock forming elements, such as Mg, Si, Ca, Fe, and lighter elements known as CHON particles (Carbon, Hydrogen, Oxygen, and Nitrogen) \citep{CometsII_dust}. Both crystalline and non crystalline silicates are present; the major fraction of non-crystalline silicates is constituted of glassy silicate grains (GEMS - glass with embedded metals and sulfides). In-situ measurements of 67P/Churyumov-Gerasimenko show many organics, and in particular, phosphorus, which is an important element in the emergence of life \citep{Altwegg_2016}. Cometary grains can be of any range of mixture from pure silicate to almost fully icy grains, and vary depending on the comet. 

Cometary nuclei are often characterized as either icy ``dirtballs" or dusty ``snowballs" \citep{CometsII_snowball}. There are no definite composition models that fit every cometary nucleus, however it can be inferred that a majority of cometary nuclei are collisionally processed rubble piles, defined as ``primordial rubble piles that have subsequently undergone collisional evolution" \citep{CometsII_snowball},``a `layered pile' model, in which the interior consists of a core overlain by a pile of randomly stacked layers" \citep{Belton_2006}, or are a collection of coalesced smaller bodies if they were formed in a streaming instability \citep{Weissman_2020}. The composition, porosity, and size distribution of particles within the inner coma, if originating directly from the inside of the nucleus, can shed light on the make-up of the cometary nucleus or if originating from the upper layers of the cometary nucleus, can shed light on the evolution of cometary nuclei in general.  

\begin{figure}
\centering
\includegraphics[width=14cm]{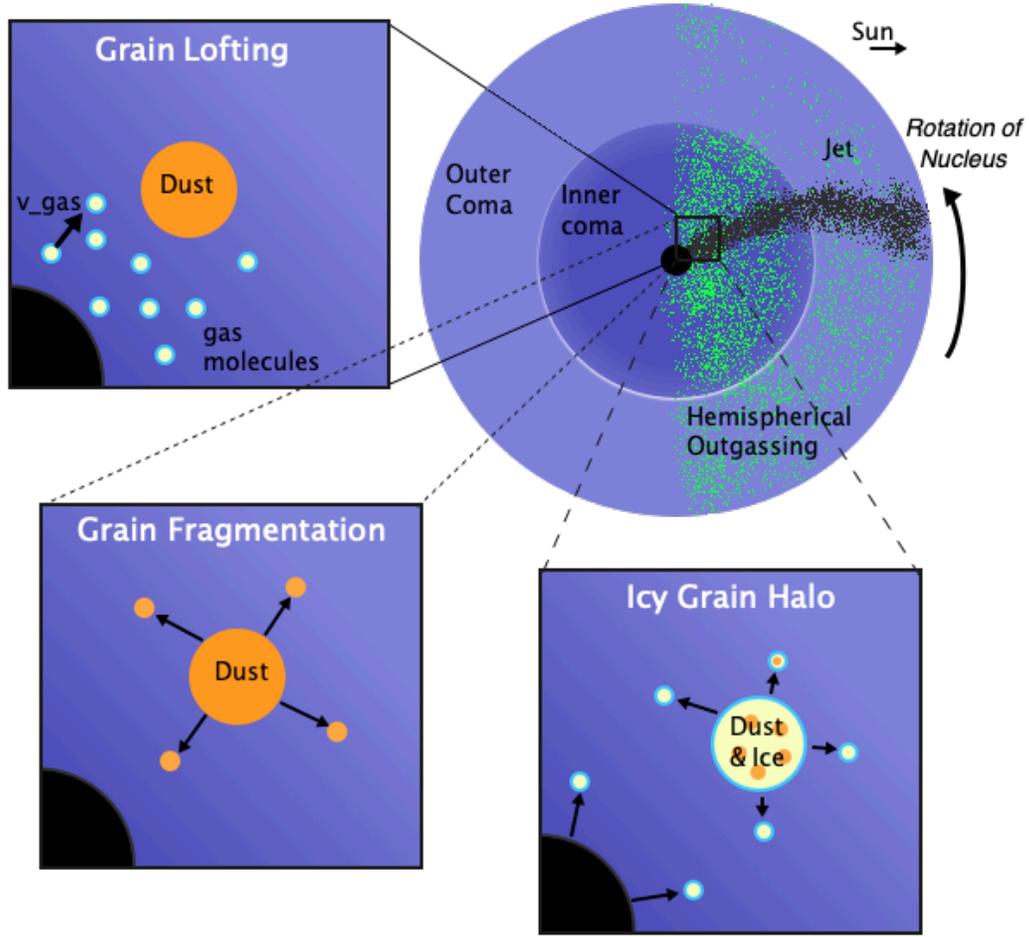}
\caption{Cometary dust undergoes many changes in the coma. The relevant processes include grain lofting, grain fragmentation, formation of icy grain halos, hemispherical outgassing, and jet features, which can be curved due to rotation of the nucleus.}
\label{dust_types}
\end{figure}

When being dragged through the coma, cometary dust particles can undergo several different processes, including grain lofting and grain fragmentation, which, if mixed with ice, creates an icy grain halo. As Figure \ref{dust_types} shows, grain lofting simply refers to gas drag forces propelling dust particles from the surface of the nucleus and bringing it along to the coma. During the grain lofting, the dust is accelerated by the gas until the gas density becomes so small that the acceleration is negligible, and the dust becomes decoupled from the gas. This region of decoupling usually defines the difference between inner and outer coma and occurs around several thousand of kilometers and primarily depends on the gas production rates. 

Grain fragmentation occurs when large particles are broken up due to their ``fluffy aggregate" structure. Grain fragmentation creates a change in observable dust reflective surface area from the simple grain lofting model. An icy grain halo is created when larger particles of dust and ice are lofted, then partially sublimate. The partial sublimation typically causes a depletion in the observable dust surface area. Both grain fragmentation and icy grain halo have been analytically replicated by \citet{Markkanen}. It is important to understand that there are many additional processes under which cometary dust grains can evolve in the inner coma, and that the apparent radial distribution depends on a number of factors that come down to whether the cross section, albedo, and velocity distribution of the grains remain unchanged or not. Understanding the composition and structure of cometary dust allows us to remotely probe each individual cometary nucleus. 

\subsection{Fountain Model \label{fountain_model}}
As explained in \S \ref{importance_dust}, it is assumed that gases leaving the nucleus accelerate away due to gas expansion among other forces. Solar radiation pressure effects are more apparent farther out from the nucleus, and following the equation of motion, the displacement caused by radiation pressure is proportional to time it is exposed to the radiation field squared. During their sublimation, gases can pick up dust particles on the surface of the cometary nucleus and entrain them, in a process called grain lofting, as they move away from the nucleus until they are decoupled from the gas. At this point, if other dust processes are negligible, the dust particles usually expand radially at a constant velocity until they are affected by solar radiation pressure. This is the idea for the Fountain Model as described by \citet{Fountain_orig}. Although there is some evolution of the radial dust outflow in all directions, with dust grains accelerating away from the sun, our paper focuses on the inner coma, where the decoupling from the gas occurs, but the radiation pressure is not yet a major factor. Essentially, the inner coma is the location where the dust transitions from being coupled with the gas to being decoupled from the gas.  A spherically expanding dust coma, that is decoupled from the gas but is essentially unaffected by radiation pressure, when projected onto the plane perpendicular to the line of sight of the observer (sky plane), will produce a brightness profile proportional to $1/\rho$, where $\rho$ is the projected distance from the nucleus on the sky plane. 

Even though the Fountain Model has been known for over a hundred years, and has often been referenced when looking at cometary dust expansion, it is important to keep in mind that the expanding gas coma, which drags the dust with it, is composed of different gases that have a range of photo-dissociative lifetimes. Photo-dissociation of molecules will cause gas accelerations that may or may not occur within the region where dust and gas are coupled. The Haser \citep{Haser} and Vectorial \citep{Festou_1981} Models represent the gas expansion in comae.  Additionally, even if the dust and gas typically decouple within thousands of kilometers from the nucleus, the actual decoupling distance will depend on the shape and size of the dust particles \citep{Ivanovski_2017} in addition to the gas production rate. Our ability to resolve the inner coma where the dust and gas might not yet be fully decoupled allows us to investigate the radial profile behavior of the dust comae, informing us of the grain environment.

\section{DATA} \label{sec:data}
\label{Data_section}

\begin{figure}[ht]
\centering
\includegraphics[width=0.48\textwidth,angle=0]{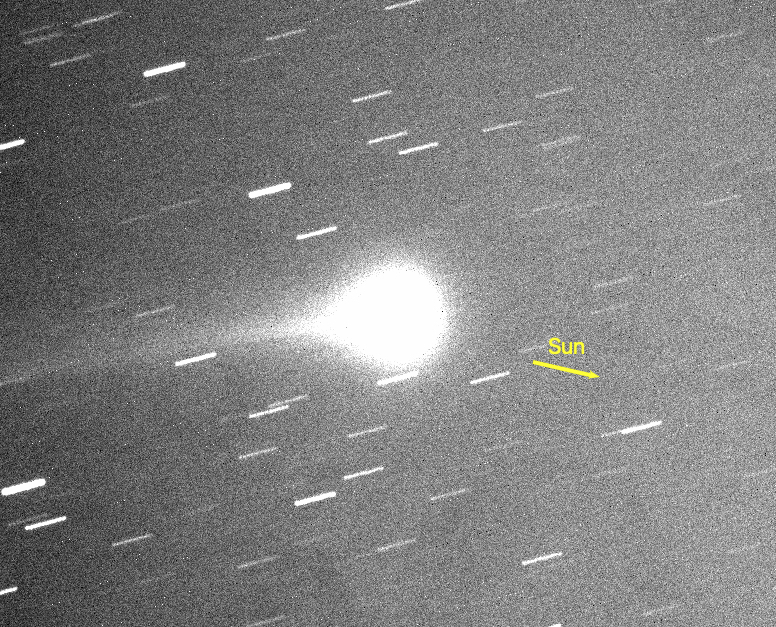}
\hspace{0.4cm}
\includegraphics[width=0.48\textwidth,angle=0]{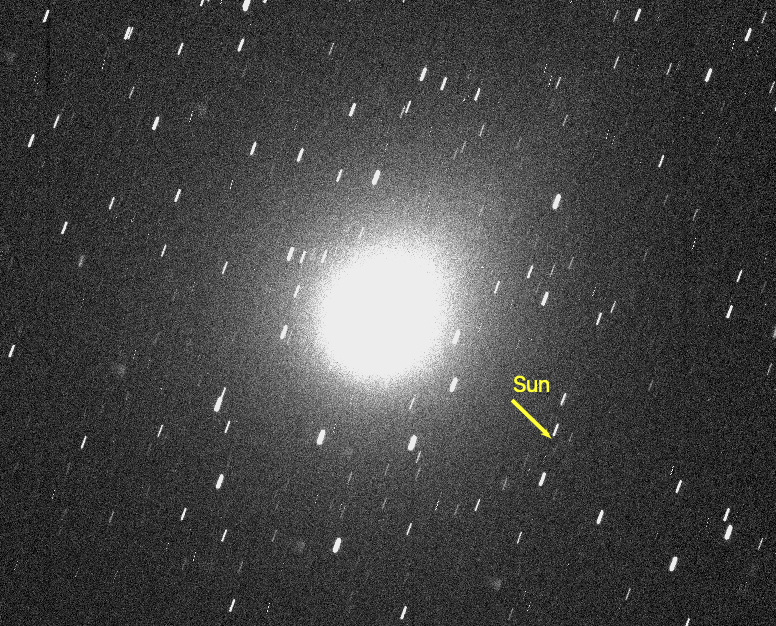}
\caption{ Images of 45P/HMP (left) on 2016-12-28 taken at the Faran telescope and 46P/Wirtanen (right) taken on 2018-12-09 taken at the ROADS telescope. Both represent the closest day to perihelion that we have for each comet. The arrow represent the projected solar direction on the plane of the sky, and represent 50,000 km  and 10,000 km, respectively. In both images, North is up, and East is to the left. \label{Pretty_image}}
\end{figure}

As seen in Table \ref{perigee_distances}, images were obtained for a large range of geocentric and heliocentric distances. The exposure times were determined to match a non-saturating nucleus, if non-sidereal tracking was available, otherwise the exposure times were optimized to avoid smearing. Images were obtained for the majority of the comets' observability in the sky for each night we had telescope time, and were on a repeating cycle through the observing campaign's set of filters.  The differences in total observing nights available were impacted by the comets' geometry, the weather, and telescopes' scheduling. It is important to note that 45P/HMP's limited observable window due to the observing geometry resulted in a smaller data set. A sample image for each comet, when closest to perihelion, is shown in Figure \ref{Pretty_image}.

\begin{sidewaystable}
\centering
\caption{The telescopes, observers,  and specific setups used in this paper are shown below. \label{4*P_data}}
\textbf{University of Arizona Observatories}\\[1.5pt]
\begin{tabular}{|c|c|c|c|c|c|c|c|c|}
\hline
\textbf{Telescope}                                                    
            & \textbf{Observer(s)}                          
            & \textbf{Camera}    
            & \textbf{Filter}       
            & \multicolumn{1}{m{1.4cm}|}{\textbf{Field of View ($arcmin^2$)}} 
            & \multicolumn{1}{m{2.4cm}|}{\textbf{Comet \hspace{1.0cm} Observed}} &\multicolumn{1}{m{1.5cm}|}{\textbf{Location}} 
            & \textbf{Citation}  
            & \multicolumn{1}{m{1.5cm}|}{\textbf{\centering Short Name}}                \\ \hline

\multicolumn{1}{|m{3.0cm}|}{1.54 m Kuiper Telescope}   
            & \multicolumn{1}{m{3.0cm}|}{Collaborative program, see Acknowledgements}     
            & \multicolumn{1}{m{2.1cm}|}{UA ITL 4k x 4k backside processed CCD} 
            & \multicolumn{1}{m{1.5cm}|}{Harris-R, HB-BC}    
            & \multicolumn{1}{m{1.4cm}|}{9.7 x 9.7}      
            & \multicolumn{1}{m{2.4cm}|}{\centering 45P/HMP 46P/Wirtanen}  
            & \multicolumn{1}{m{1.9cm}|}{Tucson,\hspace{0.5cm} Arizona, USA}    
            & \multicolumn{1}{m{1.7cm}|}{\citet{Kuiper}} 
            & \multicolumn{1}{m{1.5cm}|}{\centering 61''} \\ \hline
\end{tabular}

\vspace{6pt}

\textbf{4*P Coma Morphology Campaign Data Utilized}

 \vspace{2.5pt}
\begin{tabular}{|c|c|c|c|c|c|c|c|c|}
\hline

\multicolumn{1}{|m{3.0cm}|}{ROAD Observatory, 0.4-m (MPC G39)}              
            & \multicolumn{1}{m{3.0cm}|}{E. Bryssinck,  F.-J. Hambsch}                                                  
            & \multicolumn{1}{m{2.0cm}|}{Scientific FLI 16803 4k x 4k} 
            & R                                                             
            & 47 x 47                                                                       
            & 46P/Wirtanen    
            & \multicolumn{1}{m{1.5cm}|}{San Pedro de Atacama, Chile}                  
            & \multicolumn{1}{m{1.7cm}|}{Personal Communication}                            
            & ROAD \\ \hline

\multicolumn{1}{|m{3.0cm}|}{Osservatorio Astronomico di Savelli, 505 mm}    
            & \multicolumn{1}{m{3.0cm}|}{A. Brosio}     
            & \multicolumn{1}{m{2.1cm}|}{CCD FLI PL1001 1k x 1k camera}                
            & R                                                             
            & 21 x 21                                                                       
            & 46P/Wirtanen     
            & \multicolumn{1}{m{1.5cm}|}{Savelli, Italy}                                                
            & \multicolumn{1}{m{1.7cm}|}{\citet{Savelli}}                                   
            & Savelli   \\ \hline

\multicolumn{1}{|m{3.0cm}|}{Stazione Astronomica di Sozzago, 0.4 m}         
            & \multicolumn{1}{m{3.0cm}|}{F. Manzini, V. Oldani, P. Ochner, L. R. Bedin}     
            & \multicolumn{1}{m{2.1cm}|}{KAF-8300 CCD Moravian G3}                
            & Clear                                                         
            & 22 x 17                                                                       
            & 46P/Wirtanen       
            & \multicolumn{1}{m{1.5cm}|}{Sozzago, Italy}                                                              
            & \multicolumn{1}{m{1.7cm}|}{Personal Communication}                            
            & Sozzago        \\ \hline

\multicolumn{1}{|m{3.0cm}|}{Stazione Osservativa di Asiago Cima Ekar, 67/92-cm Schmidt telescope}
            & \multicolumn{1}{m{3.0cm}|}{F. Manzini, V. Oldani, P. Ochner, L. R. Bedin} 
            & \multicolumn{1}{m{2.1cm}|}{KAF 16803 CCD Moravian G3. 1.5k Window}        
            & Sloan R                     
            & 22 x 22                               
            & 46P/Wirtanen       
            & \multicolumn{1}{m{1.5cm}|}{Asagio, Italy}    
            & \multicolumn{1}{m{1.7cm}|}{\citet{Cima}}  
            & Asiago     \\ \hline

\multicolumn{1}{|m{3.0cm}|}{Ritchey-Chretien (D=0.4m, F/5.4)} 
            & \multicolumn{1}{m{3.0cm}|}{A. Maury, J.-B. de Vanssay, J.-G. Bosch}      
            & \multicolumn{1}{m{2.1cm}|}{ASCOM\_QHY9 CCD 1112 by 832 Camera}        
            & Clear                 
            & 29 x 22                                
            & 45P/HMP        
            & \multicolumn{1}{m{1.9cm}|}{San Pedro de Atacama, Chile} 
            & \multicolumn{1}{m{1.7cm}|}{Personal Communication}      
            & San Pedro        \\ \hline

\multicolumn{1}{|m{3.0cm}|}{Faran Observatory (D=17", F/6.8)}  
            & N. Moriya                                 
            & \multicolumn{1}{m{2.1cm}|}{FLI 16803 4k by 4k Camera}                 
            & \multicolumn{1}{m{1.5cm}|}{A BAADER UHC-S filter} 
            & 44 x 44                                
            & 45P/HMP            
            & \multicolumn{1}{m{1.5cm}|}{Mitzpe Ramon, Negev, Israel}    
            & \multicolumn{1}{m{1.7cm}|}{\citet{Netzer}}          
            & Faran    \\ \hline

\multicolumn{1}{|m{3.0cm}|}{Lulin Ovservatory (D=41cm, F/8.4)}   
            & Z.-Y. Lin   
            & \multicolumn{1}{m{2.1cm}|}{Andor Tech CCD 2k by 2k camera}  
            & R     
            & 27 x 27                                
            & 45P/HMP        
            & \multicolumn{1}{m{1.5cm}|}{Jungli City, Taiwan}        
            & \multicolumn{1}{m{1.7cm}|}{\citet{Lulin}}          
            & Lulin    \\ \hline

\multicolumn{1}{|m{3.0cm}|}{Master-Ural}  
            & \multicolumn{1}{m{3.0cm}|}{V. Krushinsky, O. Ivanova} 
            & \multicolumn{1}{m{2.1cm}|}{Apogee Alta U16 4K by 4K camera}  
            & R                     
            & \multicolumn{1}{m{1.4cm}|}{30 x 30 (cropped)} 
            & 45P/HMP   
            & \multicolumn{1}{m{1.5cm}|}{Kourovka, Russia}
            & \multicolumn{1}{m{1.7cm}|}{\citet{Ural_telescope}}   
            & Ural     \\ \hline            
\end{tabular}
\end{sidewaystable}

\subsection{Telescopes \label{telescope}}
As shown in Table \ref{4*P_data}, the data presented here were obtained from multiple telescopes. We obtained the majority of our data for both comets with the 1.54 m Kuiper Telescope of the University of Arizona Observatories maintained by Steward Observatory (see Table \ref{4*P_data}). Our data set was supplemented by data from the 4*P Coma Morphology Campaign \citep{4*P}. The 4*P Coma Morphology Campaign allowed both professional and amateur observers to submit comet images they obtained following guidelines provided. Of these submitted data sets, we have used images from the telescopes/observers provided in Table \ref{4*P_data}. It is important to note that these observations were provided by observers following a similar set of instructions, but with diverse instrumentation and filters. The images from the 4*P Coma Morphology Campaign were chosen based on necessity of time cadence and usability of the images. Occasionally, measurements made at both the 1.54m Kuiper Telescope and by the 4*P Coma Morphology Campaign were compared to assess systematic errors.

\subsection{Photometric Filters \label{Filters}}
Images were taken at the 1.54 m Kuiper telescope using the HB-BC filter (HB being a reference to Hale-Bopp and BC standing for Blue Continuum) as described in \citet{HB_Filters} and the Harris-R filter. The HB-BC narrowband filter is specially designed to isolate cometary dust in the blue continuum. The Harris-R filter, most similar to the Cousins-R filter, is a widely used broadband red filter that is dominated by the dust continuum signal, though may have some limited gas coma contamination. Having a broadband filter, such as the Harris-R, allows us to have a higher signal-to-noise ratio than narrowband filters. The HB-BC and Harris-R filters were compared on the same date to check for discrepancies in radial profile slopes, and no significant discrepancies were found. We use both interchangeably, when available, to obtain radial profile slopes, with a preference on the broadband Harris-R filter for its higher signal to noise ratio. As can be seen in Table \ref{4*P_data}, the ROAD, Savelli, Asiago, Lulin, and Ural telescopes also used a form of the R filter. In addition to these two preferred dust filters, some of the observers from the 4*P Coma Morphology Campaign used both Clear and light-pollution Clear filters. Both of these include almost all the outgassing of the comet in the visible wavelengths, from 400 nm to 700 nm, with the caveat that the light-pollution filter removes specific wavelength bands usually dominated by Earth's light pollution. Clear filters do not isolate the dust from the gas emissions, however, the implication for this will be further examined in \S \ref{45P_az_section} where it is applicable. 

\subsection{Observations \label{dates_observed}}

As described in \S \ref{telescope}, the data sets were taken from different telescopes over the globe (see Table \ref{4*P_data}). Tables \ref{45P_dates_table} \& \ref{46P_dates_table} detail observing information, including the dates used, the UT time of the middle of each image (or combined image) used, the heliocentric and geocentric distances, the solar position angle at the time of observations, the projected ranges of radial distance used in the measurements of the radial profiles, and the short name of the telescope from which that specific data were obtained. In summary, we observed 45P/HMP between 26 December, 2016 and
 15 March, 2017 and 46P/Wirtanen between 10 November, 2018 and 13 February, 2019.

\begin{table}[ht]
\centering
\caption{Dates where radial profiles were measured for 45P/HMP. The telescopes' short names are referenced in Table \ref{4*P_data}. The negative values for the heliocentric/geocentric distances represent the pre-perihelion/pre-perigee distances. \label{45P_dates_table}}
\textbf{45P/Honda-Mrkos-Pajdu\u s\'akov\'a}
\\[1.5pt]
\begin{tabular}{|ccccccc|}
\hline
\textbf{Date} & \textbf{UT Time} & \textbf{$r$\footnote{Heliocentric distance} (au)} & \textbf{$\Delta$\footnote{Geocentric distance} (au)} & \multicolumn{1}{m{2.5cm}}{\textbf{Solar Position Angle (degree)}} & \textbf{$\rho$\footnote{Projected radial distance range measured} ($\times 10^3$ km)} &\textbf{Telescope} \\ \hline
2016-12-26 & 0:45    & -0.54 & -0.82  & 258.1 & 4.7 - 79    & San Pedro \\
2016-12-28 & 16:00   & -0.54 & -0.77  & 257.4 & 3.3 - 43    & Faran             \\
2017-01-09 & 10:30   & 0.57 & -0.52  & 253.4  & 3.0 - 30    & Lulin             \\
2017-02-06 & 21:50   & 0.92 & -0.10  & 112.9  & 0.2 - 6     & Lulin             \\
2017-02-08 & 12:45   & 0.94 & -0.09  & 119.2  & 0.2 - 4     & 61"              \\
2017-02-09 & 12:00   & 0.95 & -0.09  & 116.4  & 0.2 - 7     & 61"              \\
2017-02-10 & 12:30   & 0.97 & -0.08  & 112.7  & 0.2 - 6     & 61"              \\
2017-02-16 & 10:30   & 1.05 & 0.11   & 81.3   & 0.3 - 11    & 61"              \\
2017-02-16 & 21:30   & 1.06 & 0.11   & 79.0   & 0.5 - 33    & Lulin             \\
2017-02-23 & 20:50   & 1.16 & 0.19   & 47.1   & 0.9 - 69    & Lulin             \\
2017-02-27 & 19:45   & 1.21 & 0.24   & 26.6   & 1.1 - 68    & Lulin             \\
2017-03-07 & 7:00    & 1.32 & 0.34   & 347.2  & 2.1 - 32    & 61"              \\
2017-03-15 & 16:20   & 1.43 & 0.47   & 321.4  & 5.0 - 56    & Ural             \\ \hline
\end{tabular} 
\end{table}

\begin{table}[ht]
\centering
\caption{Dates where radial profiles were measured for 46P/Wirtanen.  The telescopes' short names are referenced in Table \ref{4*P_data}. The negative values for the heliocentric/geocentric distances represent the pre-perihelion/pre-perigee distances. \label{46P_dates_table}}
\textbf{46P/Wirtanen}
\\[1.5pt]
\begin{tabular}{|ccccccc|}
\hline
\textbf{Date} & \textbf{UT Time} & \textbf{$r$\footnote{Heliocentric distance} (au)} & \textbf{$\Delta$\footnote{Geocentric distance} (au)} & \multicolumn{1}{m{2.5cm}}{\textbf{Solar Position Angle (degree)}} & \textbf{$\rho$\footnote{Projected radial distance range measured} ($\times 10^3$ km)} & \textbf{Telescope} \\ \hline
2018-11-10 & 5:20    & -1.14  & -0.22  & 197.6 & 0.9 - 135     & ROAD   \\
2018-11-13 & 8:45    & -1.13  & -0.21  & 201.1 & 0.7 - 51      & Savelli   \\
2018-11-15 & 5:35    & -1.12  & -0.20  & 202.3 & 2.0 - 129     & ROAD   \\
2018-11-23 & 5:00    & -1.09  & -0.16  & 209.0 & 1.3 - 41      & ROAD   \\
2018-11-28 & 4:45    & -1.07  & -0.13  & 212.9 & 0.5 - 110     & ROAD   \\
2018-11-30 & 4:40    & -1.07  & -0.12  & 214.6 & 0.5 - 50      & ROAD   \\
2018-12-04 & 20:15   & -1.06  & -0.10  & 219.1  & 0.3 - 59      & Sozzago   \\
2018-12-09 & 3:00    & -1.06  & -0.09  & 225.9  & 0.2 - 18      & 61"       \\
2018-12-09 & 4:00    & -1.06  & -0.09  & 226.0 & 0.9 - 85      & ROAD   \\
2018-12-21 & 5:00    & 1.06   & 0.08   & 312.5 & 0.2 - 14      & 61"       \\
2018-12-26 & 17:15   & 1.07   & 0.10   & 349.9 & 0.3 - 23      & Asiago    \\
2018-12-29 & 7:00    & 1.08   & 0.11   & 0.3   & 0.3 - 19      & 61"       \\
2019-01-05 & 6:00    & 1.10   & 0.14   & 17.4  & 0.4 - 25      & 61"       \\
2019-01-20 & 10:00   & 1.17   & 0.23   & 22.1  & 0.6 - 22      & 61"       \\
2019-01-23 & 7:45    & 1.19   & 0.25   & 20.2  & 0.6 - 23      & 61"       \\
2019-02-08 & 3:30    & 1.30   & 0.37   & 2.1   & 0.9 - 39      & 61"       \\
2019-02-13 & 7:30    & 1.34   & 0.41   & 354.9 & 1.0 - 35      & 61"      \\ \hline
\end{tabular}
\end{table}

\subsection{Data Reduction}

\begin{figure}[ht]
\centering
\includegraphics[width=15.0cm, angle=0]{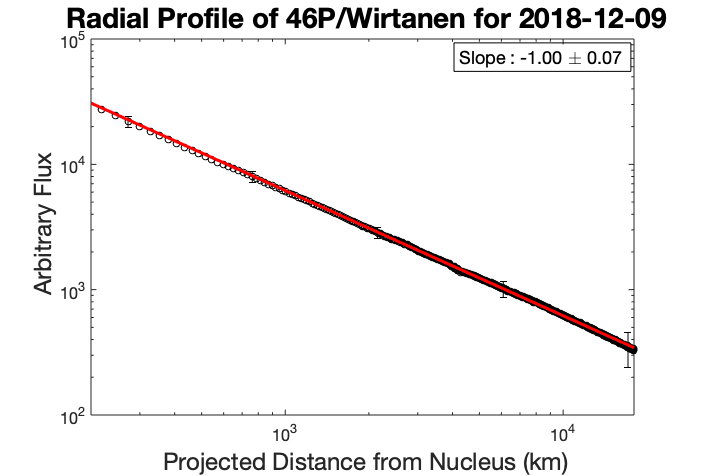}
\caption{Azimuthally medianed radial profile for 46P/Wirtanen for 2018-12-09 showing five representative error bars for total flux measurements. As visible, the error bars in individual flux measurements are not as important for fitting the slope. The small bump at about $4 \times 10^3$ in this radial profile is caused by the CCD imperfections that could not be removed in the reduction process. The red line shows the fit. \label{radial_profile_error}}
\end{figure}

Once data were obtained at each telescope, they all went through the basic reduction process that included bias subtraction, flat-fielding, and dark subtraction, when necessary. Best efforts were made to measure only the residual dust signal by removing the background flux. Although the lunar illumination was variable over our observing range, it only affected our total flux background, and not the shape of the residual cometary signal. Radial profiles were created by measuring the azimuthal median flux as a function of projected distance ($\rho$) from the nucleus (e.g., Fig. \ref{radial_profile_error}). Radial profile slopes were measured both for 30$^{\circ}$ wedges for all azimuths, and for the full 360$^{\circ}$. Linear fits were applied and the slopes were recorded. Although the uncertainty is most significant near the edges of the coma, where the overall flux is lower, the errors in the individual data points are a combination of errors in the flat-fielding, background removal, and photometric noise. The error in the background removal dominates at greater distances from the nucleus, while the photometric noise dominates closer to the nucleus as seen in figure \ref{radial_profile_error}, which can then affect the radial profile slope measured. Thus the errors of the slopes of the radial profiles are also a combination of errors in the flat-fielding, background removal, and photometric noise.

\section{RESULTS} \label{sec:results}
\begin{figure}
\centering
\begin{minipage}{.48\textwidth}
  \centering
    \includegraphics[width=\textwidth]{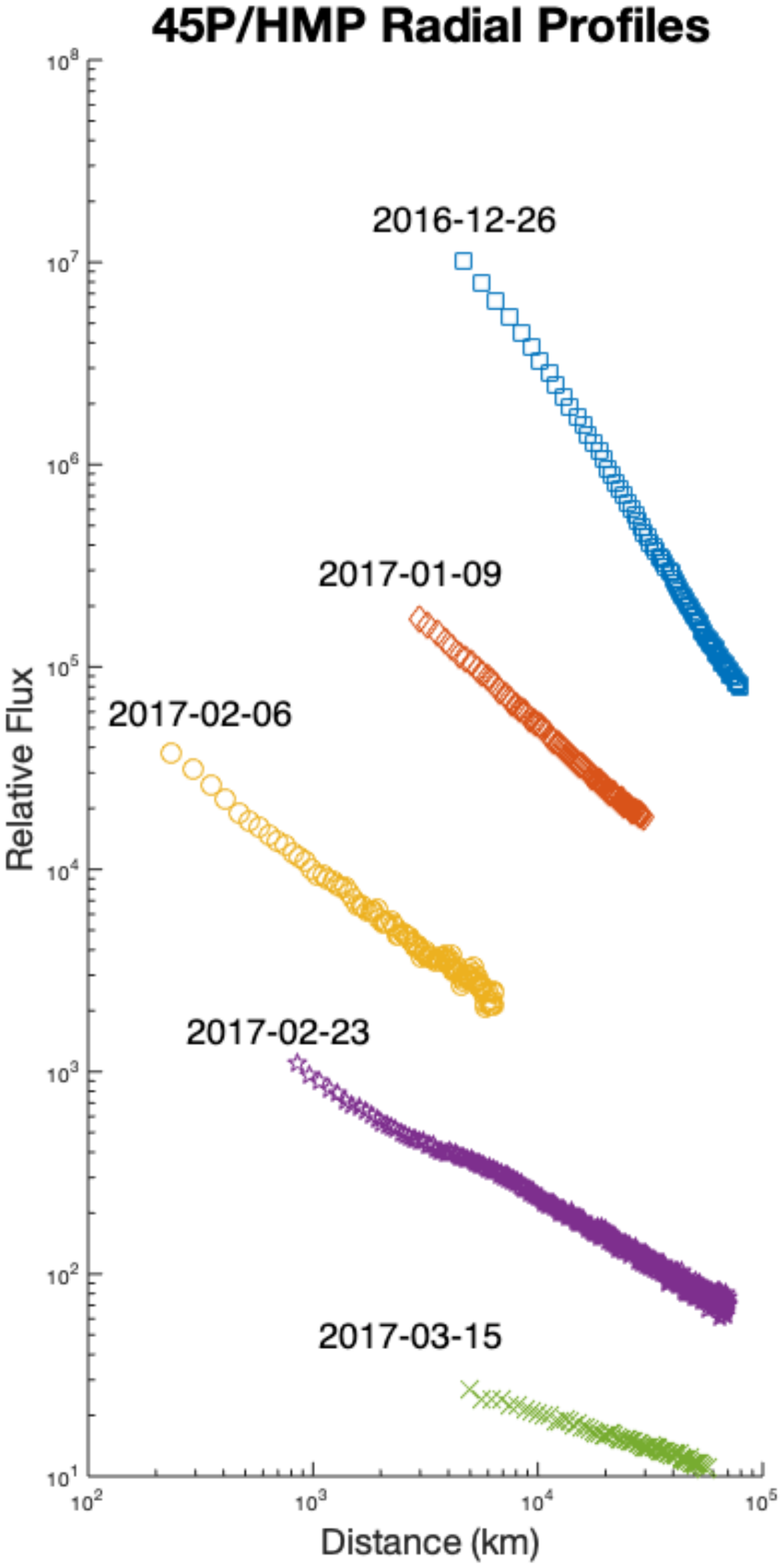}
\end{minipage}%
\begin{minipage}{.48\textwidth}
  \centering
    \includegraphics[width=0.95\textwidth]{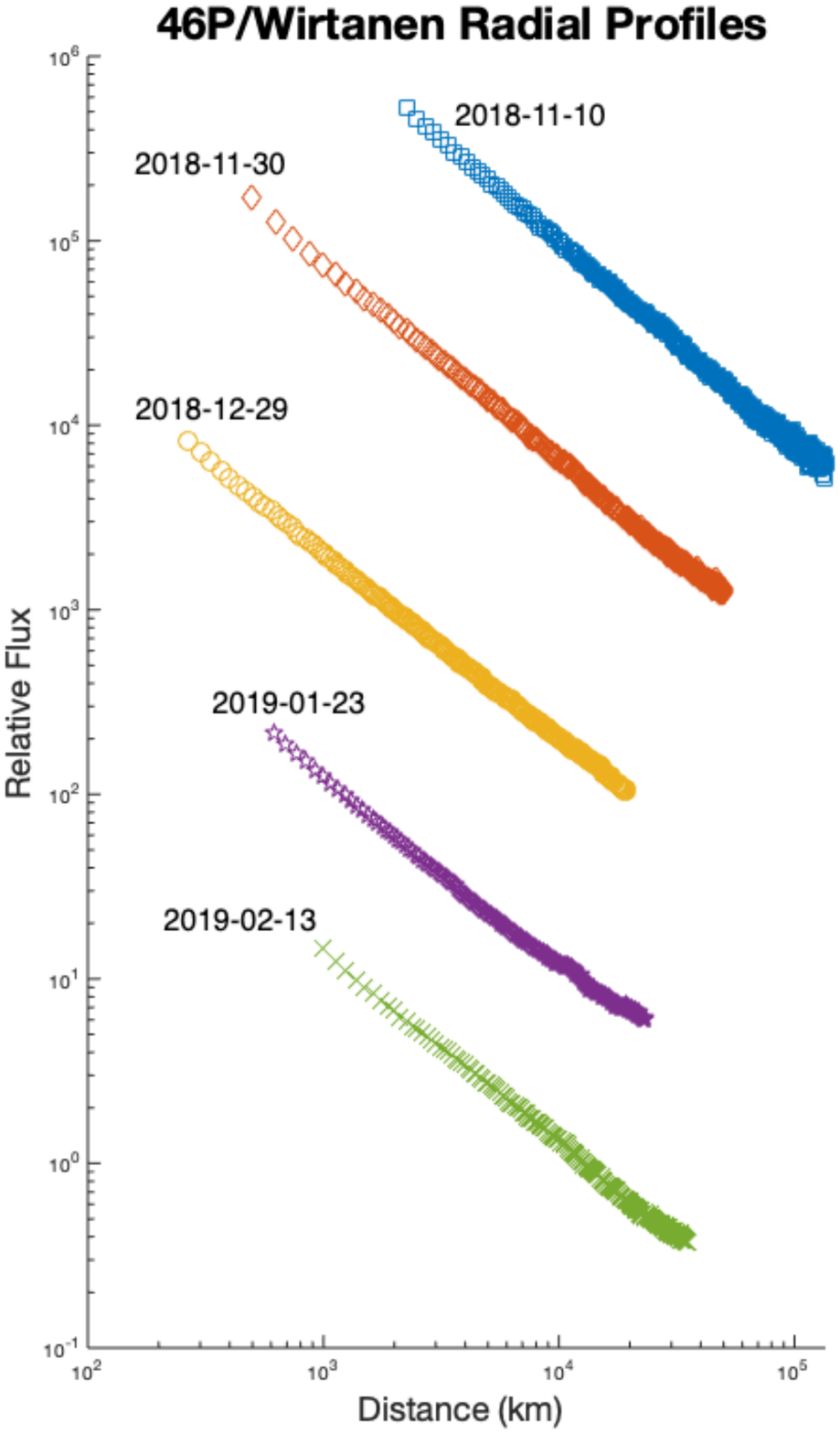}
\end{minipage}
    \caption{Radial Profiles for five dates for comet 45P/HMP (left) showing the shallowing of the slope over the observing range and comet 46P/Wirtanen (right) showing a constant slope over the observing range.}
    \label{45P_rad_prof_multiple}
    \label{46P_rad_prof_multiple}
\end{figure}

In this section, the flux due to dust continuum as a function of the projected distance from the nucleus, $\rho$, will be represented in a log-log plot and will be referred to as radial profiles. Figure \ref{45P_rad_prof_multiple} shows multiple radial profiles over the range of our observed data. The data were plotted to preserve the shape of each curve, but not the actual flux measurements, for clarity and comparison. 

\subsection{Comet 45P/HMP}
 \label{45P_az_section}
\begin{figure}[ht]
\centering
\includegraphics[width=12cm]{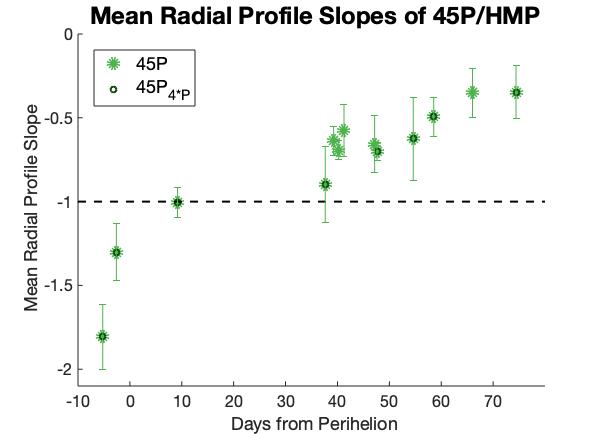}
\caption{Azimuthally medianed radial profile slopes of 45P/HMP versus days from perihelion showing a clear shallowing in the slope. The steady state fountain model is represented with a dashed line. Data points with a green circle represent those images taken by the 4*P Coma Morphology Campaign \citep{4*P}. }
\label{45P_radial_slopes}
\end{figure}

Azimuthally medianed radial profile slopes of 45P/HMP taken for the dates in Table \ref{45P_dates_table} are shown in Figure \ref{45P_radial_slopes}. Radial profiles slopes were measured from two different observatories on 2017-02-16 to confirm the relationship between 4*P Coma Morphology Campaign data and the 1.54m Kuiper Telescope data (Fig. \ref{45P_radial_slopes}, 47 days after perihelion). For each date, radial profile slopes were also measured in azimuthal directions every 30$^{\circ}$ wedge starting from north.

Figure \ref{45P_radial_slopes} shows a continuous increase in the radial profile slopes from pre-perihelion through perihelion to post-perihelion. The slope goes from much steeper than the Fountain Model expectation (\S \ref{fountain_model}) at $-1.81 \pm 0.20$ at 5.24 days pre-perihelion to much shallower at $-0.35 \pm 0.16$ at 74.41 days post-perihelion. It is important to note that the first two data points are taken when the comet is 0.82 au from Earth and with clear filters rather than an R filter. The geocentric distance of these data points causes the inner coma to be indistinguishable from the outer coma. Additionally, since we are not specifically isolating the dust, we expect gas contamination. However, the major gas contaminants, typically C$_2$, C$_3$ and CN, are chemical daughter or grand-daughter products and therefore create a radial profile slope shallower than $-1$, especially close to the nucleus, then become steeper as they dissociate. Hence, the measured slopes shown are typically upper estimates with the caveat that they are partially contaminated by the gas.

\begin{figure}[ht]
\centering
\includegraphics[width=16cm, angle=0]{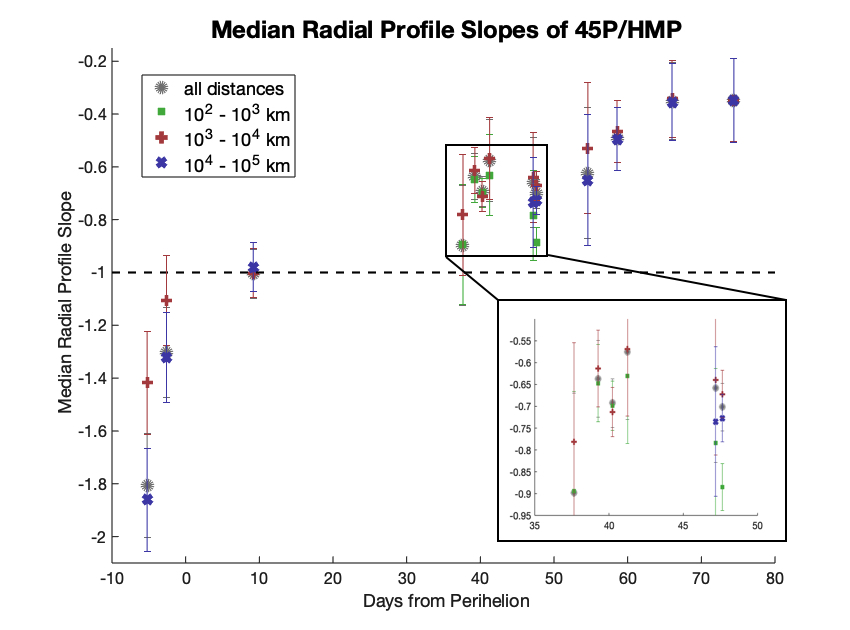}
\caption{Azimuthally medianed radial profiles binned by distance from the nucleus. The zoom in panel shows in greater details the days close to perigee where it might be otherwise hard to see the difference between the distance bins. ``All distances'' includes data points that may be outside the specific bins present for that specific day but where there were insufficient additional data to obtain statistically significant result for the additional distance bin.  \label{45P_All_slopes_binned}}
\end{figure}

Additionally, from Figure \ref{45P_rad_prof_multiple}, it is possible to notice that a potentially different radial profile slope could be measured for different distances from the nucleus. We have created distance bins in which we measured the radial profile, when it was feasible to do so for our images. The distance from the nucleus bins were as follows $10^2$ km -$10^3$ km, $10^3$ km - $10^4$ km, and $10^4$ km - $10^5$ km . Figure \ref{45P_All_slopes_binned} shows the results from the azimuthally median radial profile slopes binned by distances. Data are not available for each bin on each day primarily due to different geocentric distances and different image sizes. See Table \ref{45P_dates_table} for specific distances measured. For the six dates for which we have measurement in the $10^2$ km -$10^3$ km bin, the median profile slope is on average steeper than for the $10^3$ km - $10^4$ km distance bin by 0.09 $\pm$ 0.08. Contrarily, the radial profile slope is almost always shallower for the $10^3$ km - $10^4$ km distance bin than it is for the $10^4$ km - $10^5$ km distance bin by an average of 0.33 $\pm$ 0.16 for pre-perihelion and 0.04 $\pm$ 0.05 post-perihelion. It is important to note however that the differences in local slopes binned by distance compared with the unbinned profile slopes, as shown in Figure \ref{45P_All_slopes_binned} are comparable to the 1$\sigma$ uncertainty and must be considered to be only marginally significant.

We also analyzed the azimuthal variation of the radial profile slope. Specifically, we can analyze the slope at different azimuthal directions from the projected solar position angle. We plot the difference between the median slope and the slope of a 30$^{\circ}$ wedge as a function of the offset from the solar position angle. The offset is calculated as $PA - PA_{\odot}$, where $PA$ is the position angle measured from north counterclockwise, and the $PA_{\odot}$ is the skyplane projected solar position angle as given by JPL Horizons On-Line Ephemeris System \citep{Horizon}.

\begin{figure}[ht]
\centering
\includegraphics[width=14cm, angle=270]{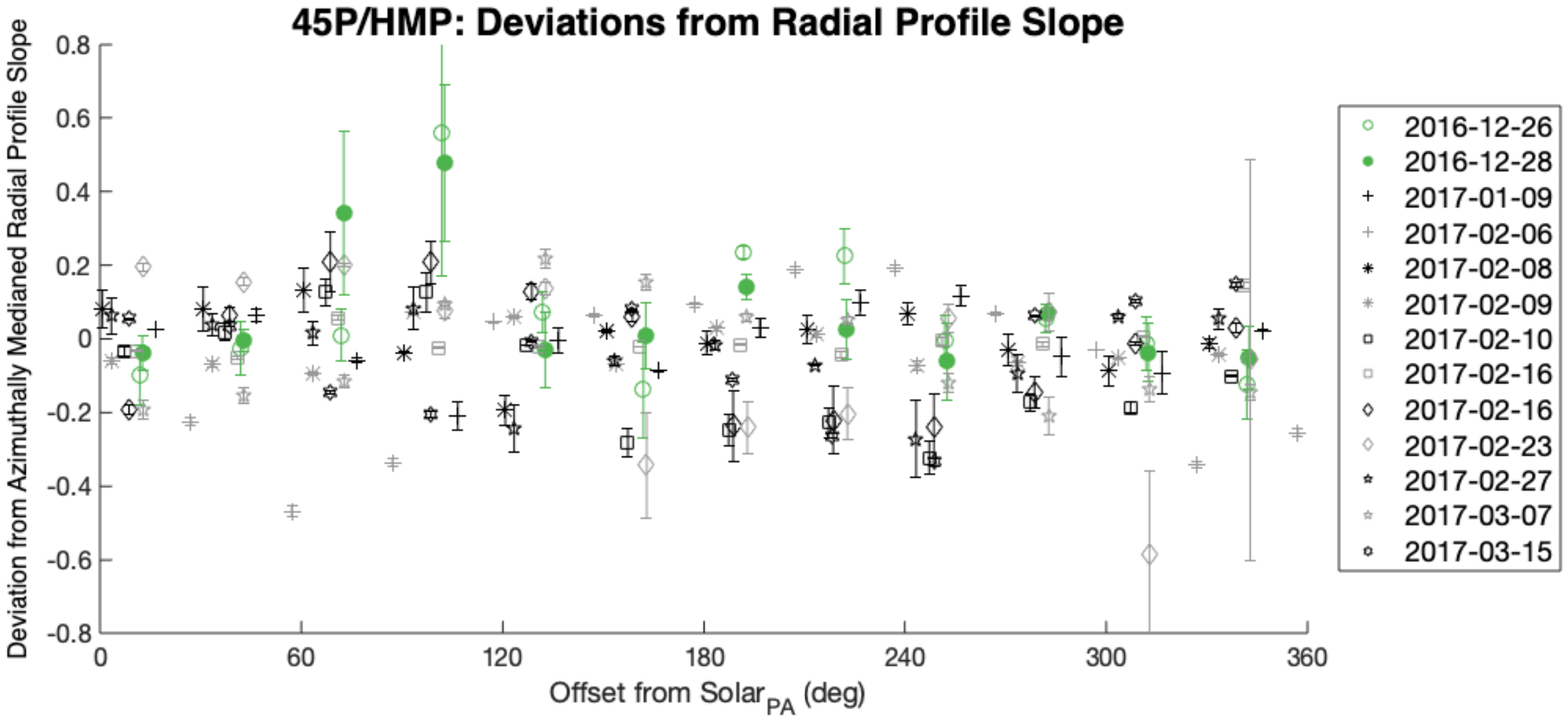}
\caption{Deviations from azimuthally median radial profile slope for 45P/HMP versus the offset from the solar position angle ( $PA-PA_{\odot}$) measure every 30$^{\circ}$ centered on North. The green markers are pre-perihelion while the gray markers are the post perihelion measurements. The error bars here represent the error in slope based upon the error on the background measurements. \label{45P_az_deviation}}
\end{figure}

Figure \ref{45P_az_deviation} shows the deviation from the median radial profile slopes versus solar position angle offset. When we look at specific dates, there may be some slight azimuthal discrepancies, but in general, there is no correlation with $PA_{\odot}$ offset. This mostly constant deviation from the median radial profile slopes throughout different azimuthal directions suggests that our radial profile slopes measured are a good representation of the radial profile slope each day and are not significantly influenced by jet features, the dust tail, nor radiation pressure. 

\begin{figure}[ht]
\centering
\includegraphics[width=0.35\textwidth,angle=270]{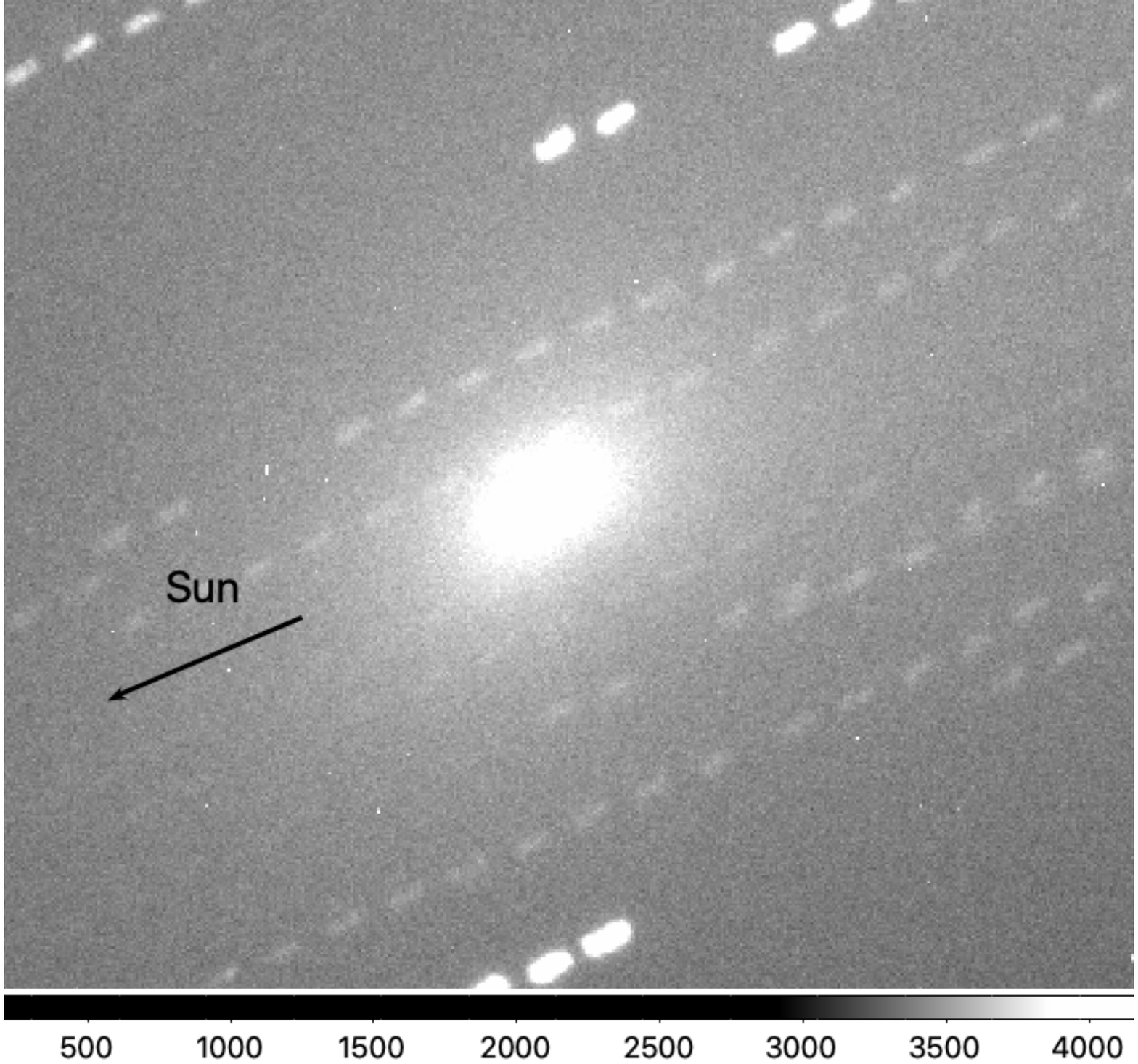}
\includegraphics[width=0.35\textwidth,angle=270]{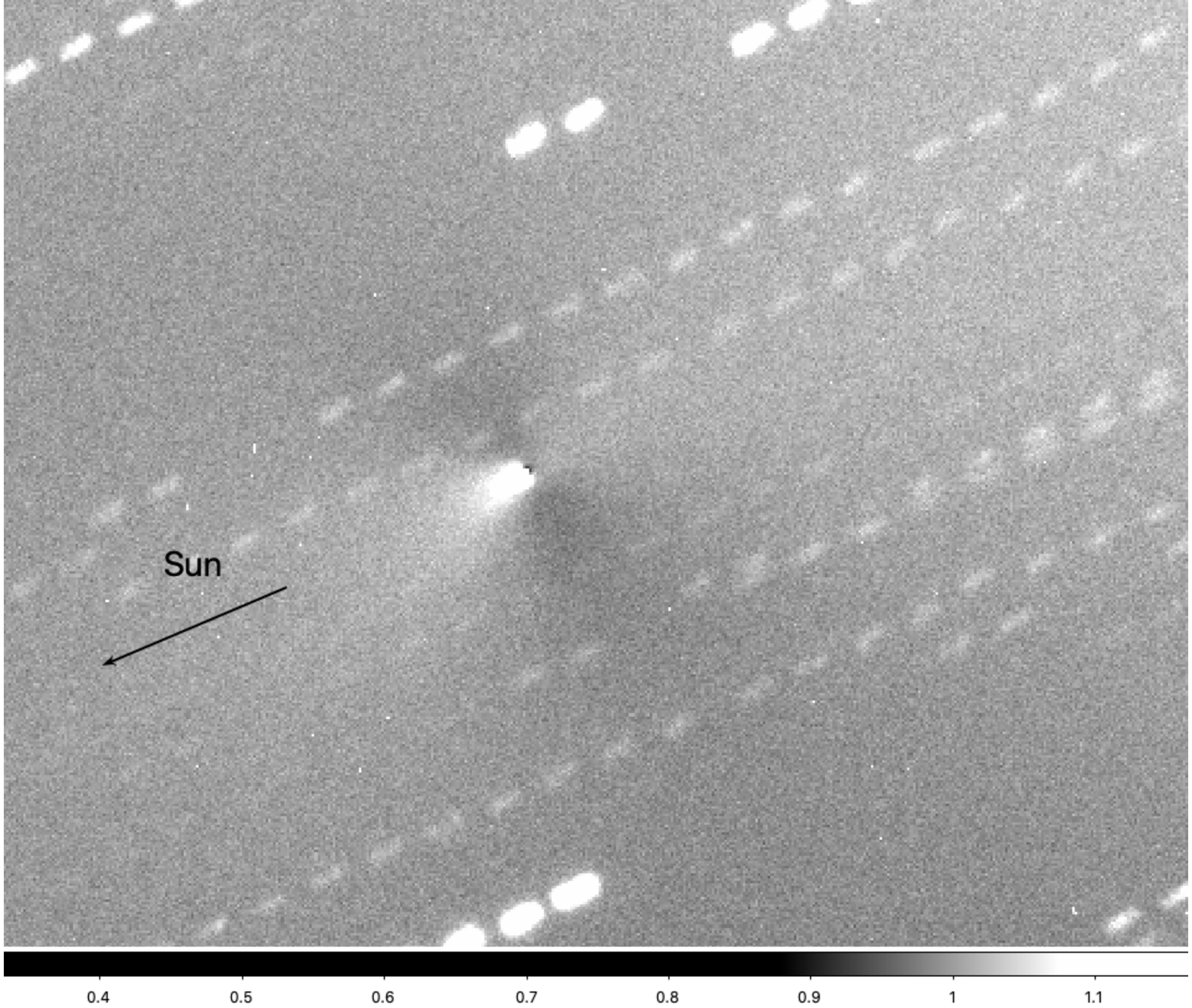}
\caption{ Left: Standard reduced image of 45P/HMP on 2017-02-10. Right: Same image divided by a circularly symmetric radial profile model of the azimuthally medianed radial profile slope measured, showing a sunward feature. In both images, the arrow length represents 2000 km, and North is up, East is to the left. \label{45P_subtraction}}
\end{figure}

To confirm the presence, or lack thereof, of significant radial features, we model a dust coma numerically as a circularly symmetric radial profile of the median measured radial profile slope for a specific image. We then divide the original observed image by the modeled dust coma to reveal underlying features, similar to \citet{nalin_enhancement}. This allows us to compare the variations in the radial profile slopes with the physical features we are able to observe. Figure \ref{45P_subtraction} shows the standard reduced image and the same image divided by a circularly symmetric radial profile. There appears to be a sunward feature visible in this enhancement technique. Its significance is discussed in more detail in \S \ref{Solar_PA_correlation_section}.

\subsection{Comet 46P/Wirtanen \label{46P_radial_profile}}

\begin{figure}[ht]
\centering
\includegraphics[width=12cm]{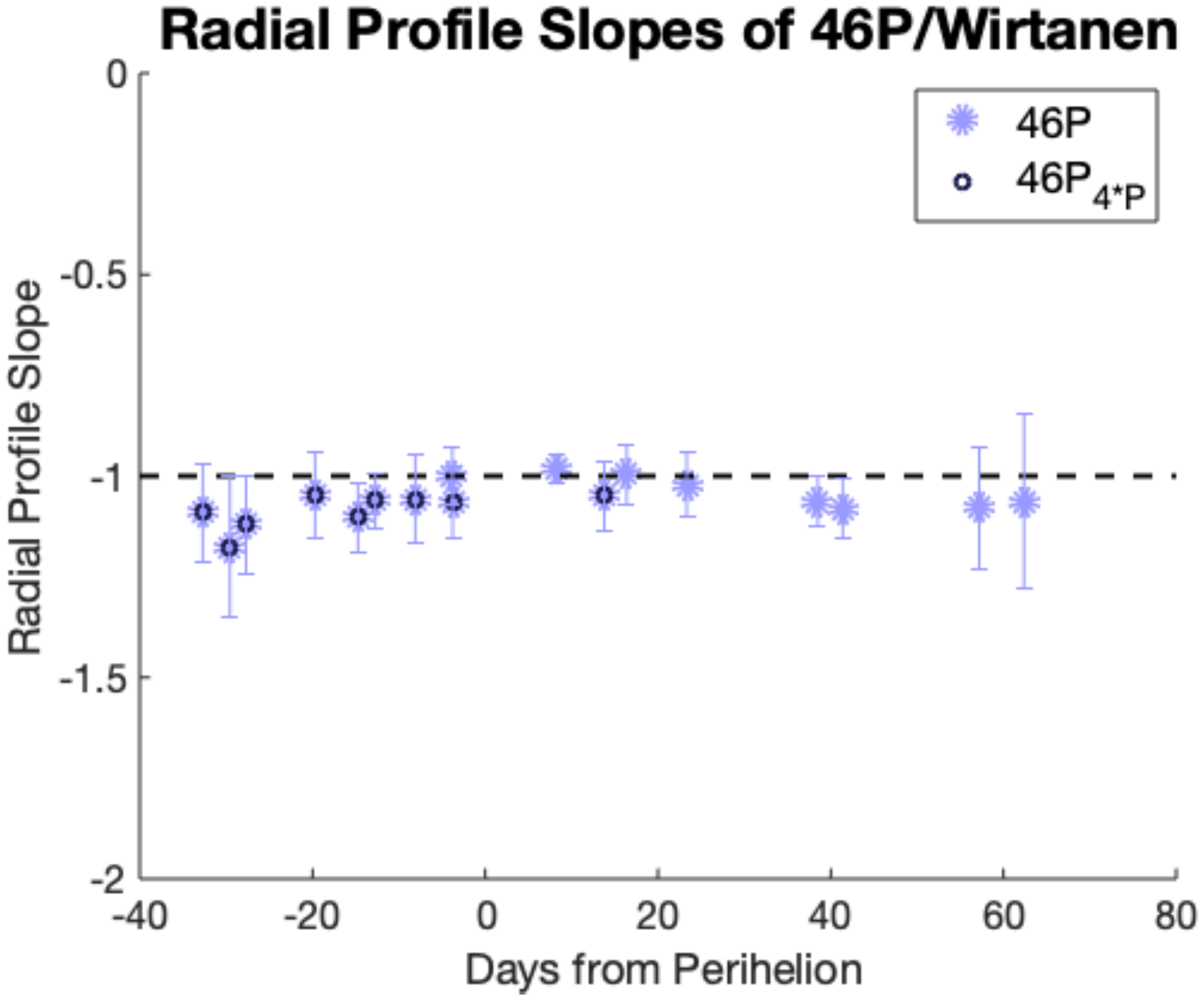}
\caption{Azimuthally median radial profile slopes of 46P/Wirtanen versus days from perihelion showing a fairly constant slope. The steady-state fountain model is represented with a dashed line. Data points with a blue circle represent those images taken by the 4*P Coma Morphology Campaign \citep{4*P}.}
\label{46P_radial_slopes}
\end{figure}

The radial profile slopes of 46P/Wirtanen were measured on 16 different nights as listed in Table \ref{46P_dates_table} and are presented in Figure \ref{46P_radial_slopes}. Eight of the nights were taken with the 1.54 m Kuiper Telescope while the rest were taken as part of the 4*P Coma Morphology Campaign. Data for the night of 2018-12-09 were analyzed and a radial profile slope was measured from both the 1.54m Kuiper Telescope and the BRIXIIS telescope as a comparison between our data and the 4*P Coma Morphology Campaign data. As for 45P/HMP, we also measured the radial profile slopes at different azimuthal directions every 30$^{\circ}$ starting from North.

Figure \ref{46P_radial_slopes} shows the radial profile slopes for 46P/Wirtanen as a function of days from perihelion. For 46P/Wirtanen, the radial slopes do not seem to change greatly over the apparition. If we take the weighted average of all the slopes, we obtained a slope of $-1.05 \pm 0.05$, which is very close to the slope of $-1$ that we would expect from a spherically expanding Fountain Model. Measurements were not done at different distances from the nucleus because of the consistency of the slopes as seen in Figure \ref{46P_rad_prof_multiple}.

\label{Slope_v_PA_46P}

\begin{figure}[ht]
\centering
\includegraphics[width=14cm, angle=270]{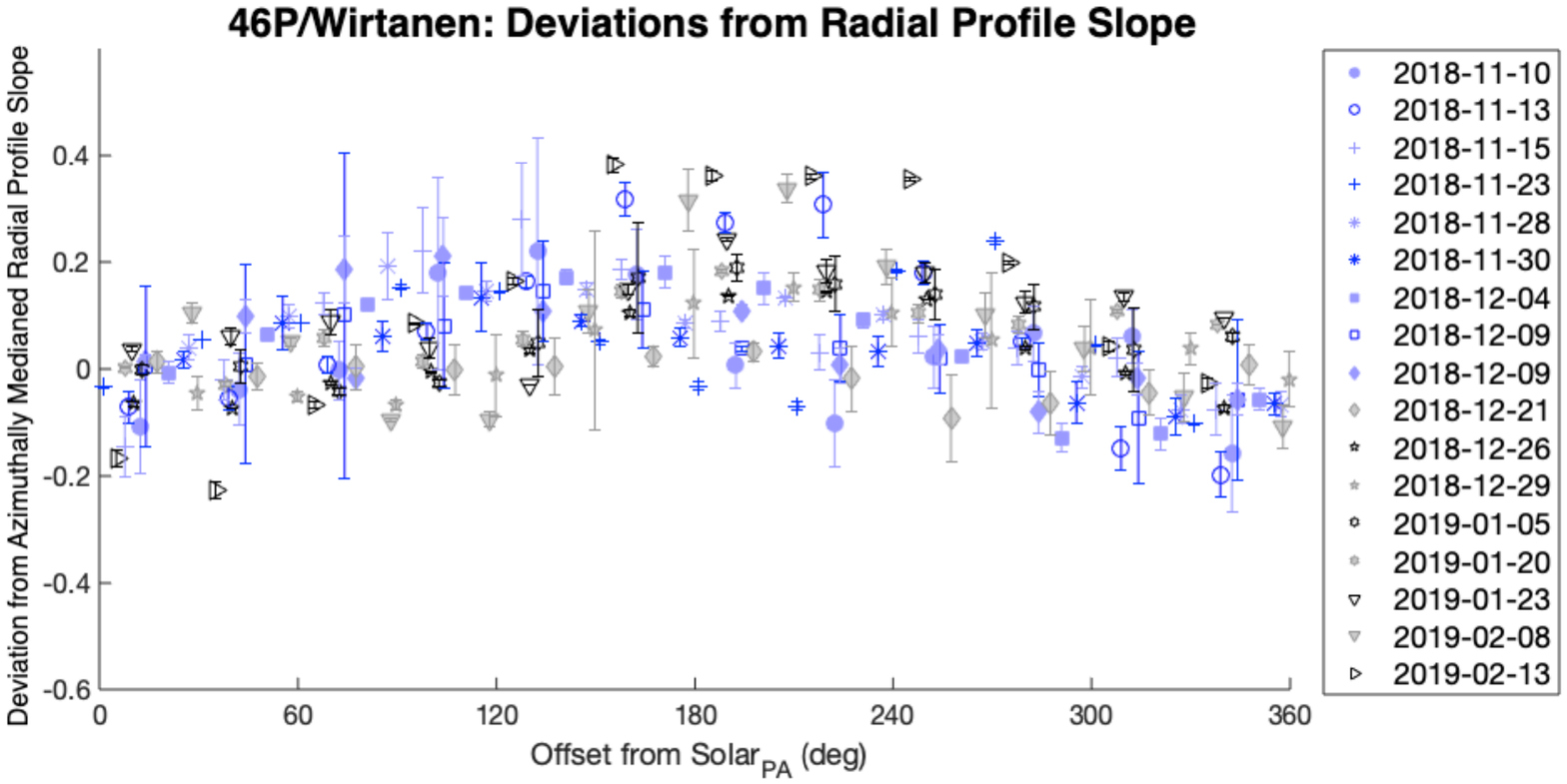}
\caption{Deviations from median radial profile slope for 46P/Wirtanen versus the offset from the solar position angle ($PA-PA_{\odot}$) measured in 30$^{\circ}$ wedges starting from North. The blue markers are pre-perihelion while the gray markers are the post-perihelion measurements. The error bars here represent the error in slope based upon the error of the background measurements.\label{46P_az_deviation} }
\end{figure}

Figure \ref{46P_az_deviation} shows the deviations from the radial profile slopes versus offset from the projected solar position angle (\S \ref{45P_az_section}). Comet 46P/Wirtanen seems to have a clear trend in the deviation from the median radial profile slopes centered at 180$^{\circ}$, counterclockwise from the solar position angle (i.e., the anti-sunward direction).  It appears that the radial profile is shallower, by approximately 0.1 in the anti-sunward direction. We describe in \S \ref{Solar_PA_correlation_section} a possible reason for this phenomenon.

\section{DISCUSSION} \label{sec:analysis}

Prior to comparing our results with the relevant literature, it is important to reiterate the specific observing geometries corresponding to our observations. Because of the very close geocentric distances of both comets, we were able to resolve the inner comae in our observations of these comets, something rarely achievable except by spacecraft (for the ranges of $\rho$ probed, see Tables \ref{45P_dates_table} and \ref{46P_dates_table}). 

Our measurements show comet 45P/HMP changing from a radial profile slope of -1.81 $\pm$ 0.20 to -0.35 $\pm$ 0.16 from pre-perihelion to post-perihelion. Furthermore, 45P/HMP's radial profile slope seems to be changing, though remains within a 1$\sigma$ error, with different binned distances from the nucleus of the comet. These are unusual behaviors which we try to characterize below. For comet 46P/Wirtanen, we obtained a fairly constant radial profile slope of $-1.05 \pm 0.05$ over the span of 95 days, which suggests a steady state coma following the Fountain Model \citep{nalin_enhancement}. However, comet 46P/Wirtanen does appear to have an azimuthal dependence. \citet{Coulson} measured the continuum emission radial profile at $850 \mu m$ for comet 46P/Wirtanen, and found a slope of -1 as close as at least 630 km from the nucleus for the dates of 2018 Dec. 14-20. This shows that our measurements are consistent with independently obtained measurements at different wavelengths representing larger grains.

\subsection{Dust Fragmentation of 45P/HMP \label{grain_fragmentation_45P}}

Our results highlight the importance of closely analyzing the dust radial profile slopes and understanding the behavior of the dust and the dust-and-gas coupling properties (see \S \ref{sec:intro}). When analyzing the radial profile slopes, it is important to consider water production rate, dust grain sizes, and grain cohesiveness. As water production rates increase, both the dust production rate and potentially the dust particle sizes in the coma increase. Larger, fluffier grains are more likely to fragment early due to gas pressures, while small cohesive grains will fragment less easily. In the case of a fragmenting fluffier grain, we can expect a shallowing of the dust profile slope, while a cohesive, unfragmenting grain, would tend to follow a slope of $-1$. As clearly visible in Figure \ref{45P_radial_slopes}, comet 45P/HMP has a radial profile slope going from very steep to very shallow. Factors that would cause a deviation from the $1/\rho$ dust radial profile  from the Fountain Model include 1) asymmetry in the gas production distributions (e.g. jet features), 2) multiple velocity distributions for the dust, 3) variable grain-size distributions, 4) variable mass loss rates, 5) radiation pressure effects, especially farther out from the nucleus, and 6) the evolution of dust with radial distances from the nucleus. This last point is particularly important because of processes such as grain fragmentation and sublimation in an icy grain halo (see \S \ref{importance_dust}), which has an important implication for 45P/HMP.

\citet{Combi} show that 45P/HMP is reasonably symmetric in the gas production rates pre- and post-perihelion, and if we assume that dust has to be driven by some type of gas behavior, we find that our asymmetry pre- and post-perihelion in Figure \ref{45P_radial_slopes} to be of importance.  Dust velocity dispersion, variable grain-size distribution, and variable mass loss are all beyond the scope of this paper, and therefore can only be speculated on when trying to find a possible explanation. 

Equation 4 from \citet{Bea_2013} provides a method for calculating the turn back distance of dust grains in the coma of comets. For comet 45P/HMP specifically, we can use our current knowledge to calculate a best estimate of this distance. The utilized parameters are as follows: 1) the grain outflow velocity is a minimum of 15 m$s^{-1}$ for centimeter-sized grains (Howell, personal communication), 2) the heliocentric distances are mentioned in Table \ref{45P_dates_table}, 3) the  range of solar phase angle are 17.1$^{\circ}$ at 0.54au pre-perihelion and 167.7$^{\circ}$ at 1.43au post-perihelion \citep{Horizon}, 4) the ratio of radiation pressure to gravitational pressure being $\sim$0.01 for grains of a few microns in size \citep{Burns_1979}, and 5) we can assume a skyplane angle of 45$^{\circ}$ as an average since we have no further knowledge of the subject. With these parameters, we obtain a turn back distance ranging from 265 - 6572 kms. This is a lower estimate from a minimum velocity, and distances may be much higher. Additionally, factors such as total gas production rates, grain sizes, and albedo of the dust were not accounted for specifically.  While some grains are likely to reach a turn back distance, on average, since our velocity is physically going to be higher for smaller grains, there is a high likelihood that a majority of our grains will not turn back within our field of view, and they will not have a major role in the radial profile.  

Thus, if we assume the other factors are in play, we can try to understand the behavior we see in Figure \ref{45P_radial_slopes} by first understanding the behavior of an optically thin symmetric dust coma. As explained in \citet{Jewitt_1987}, for an optically thin symmetric coma generated by a constant source with constant velocity dust grains that have retained their scattering properties, the radial surface brightness profile is given by $B(\rho)=K/\rho$ where $K$ is a constant. A simple way to look at the behavior of dust in the inner coma is to represent radial brightness as a function of projected distance in a log-log plot. This allows the slope of the radial profile to become 

\begin{equation}
    m=\frac{d log B(\rho)}{d log \rho},
    \label{slope_equation}
\end{equation}
where the steady-state symmetrical case described above would lead to $m=-1$ \citep{Jewitt_1987}. Deviations from $m=-1$, such as a grain fragmentation scenario would make the slope shallower (less negative) while an icy grain halo scenario would cause a steeper slope (more negative). In the case of grain fragmentation, we would increase the amount of reflected sunlight progressively farther from the nucleus by increasing the net grain cross-section. This increases the brightness of the coma farther away from the nucleus. Conversely, in the case of icy grains, as we go farther from the nucleus, we would be losing total mass of dust, thus effective cross-section. The total albedo of the grains would also be reduced, as the composition changes from icier material (high albedo) to dustier material (low albedo). This decreases the overall brightness of the coma farther away from the nucleus.  Multiple processes can occur simultaneously in the dust expansion (e.g.: fragmentation, sublimation, and albedo changes) with the net effect on the radial profile slope being dependent on which effect dominates the observed properties.

One possibility to explain the reduction in radial profile slope for 45P/HMP would be gradual changes to the types of dust grains leaving the comet, essentially creating a variable grain-size distribution. We can imagine large icy grains originally leaving the comet as it approaches perihelion, then, those having been depleted, and the outgassing becoming weaker as the comet moves away from the sun, only the small non-icy grains being released and continuing to fragment. The extremely steep radial profile slope pre-perihelion would suggest that a very rapid process existed to reduce the total flux from dust grains as they moved farther from the nucleus.

Dust fragmentation is a possibility, especially if much larger icier grains are present pre-perihelion than post-perihelion. If the slope continues to become shallower, it suggests that the grains are becoming more friable, but, however, may contain less ice. Furthermore, the shallowing of the radial profile slope with distance for pre-perihelion of 0.32 $\pm$ 0.16 from $10^3$ km - $10^4$ km to $10^4$ km - $10^5$ km suggest an even more complex evolution of the grain in the coma such as a mix of icy and dusty grains, or simply extremely icy grains that fragment into smaller still icy grains. The post-perihelion change in radial profile slope in these distance bins seem minimal enough that such a complex process may have stopped by then and possibly only non-icy grain may be at play during post perihelion. 

\subsection{Acceleration of Dust Grains Due to Outgassing}

Before the dust grains reach their terminal velocity and are still being accelerated by gas pressures, we expect a steep brightness profile near the nucleus. A rough estimate is made by \citet{Jewitt_1987} based on mostly water production at 1 au from the sun to calculate an acceleration zone, $X_a$: 
\begin{equation}
    3r_n \leq X_a \leq 30r_n,
\end{equation}
where $r_n$ is the nucleus radius. From radar images, an estimate of 45P/HMP's diameter is 1.3 km \citep{Lejoly} while 46P/Wirtanen is approximately 1.4 km in diameter (personal communication, Ellen Howell) giving us 2 km $\lesssim X_a \lesssim$ 20 km  for both comets. In our observations, our best resolution at the optimal geometry gives us 25 km/pixel, but unable to resolve the rapid acceleration of dust by the sublimating ices in the nucleus.

The issues of acceleration depend on total gas production rate, grain size, and albedo, which is where the order of magnitude comes into play. This is only a rough estimate and it is possible to have residual effects of that acceleration at distances of 100 km. The slight steepening of the slope of 0.09 $\pm$ 0.08 seen near perigee (about 40-50 days post-perihelion) in Figure \ref{45P_All_slopes_binned} could be explained by the dust being accelerated by gases leaving the nucleus.

Since dust and gas are coupled near the nucleus, the behavior of gas itself can also affect the dust radial profile. In \citet{Combi}, power laws for water production of both comets are fitted to their data. Though not symmetric about perihelion for either comet, there does not seem to be any odd behavior that could cause the radial profile slope measurements we observe for 45P/HMP (Fig. \ref{45P_radial_slopes}), such as sudden brightening, nucleus fragmentation, etc. For 45P/HMP,  \citet{Dello_Russo} measures a post-perihelion water production rate of $Q(H_2O)=(2.81 \pm 0.25) \times 10^{27}[r^{-3.83 \pm 0.18}]$ molecules s$^{-1}$; it is also stated that there may be significant variability in $H_2O$ production on time scales of days and even hours. This could cause slope changes on small time frames, but not a global shallowing trend as visible in our data. When looking at production rates of other gas molecules (OH, CN, C$_3$, and C$_2$), all appear to have a similar slightly decreasing trend from 40-70 days post-perihelion \citep{Moulane}. Therefore, it does not appear that the gas behavior alone can explain our increase in radial profile slopes for 45P/HMP.

\subsection{Dust Grain Behavior's Effects on Radial Profiles}

The size, density, composition, outflow velocity, and overall behavior of dust grains can greatly affect the radial profiles, both globally and in specific azimuthal directions on a specific date. All of these parameters are difficult to constrain; however, we have some indication of grain sizes and velocities. From the Arecibo Observatory radar measurements of both 45P/HMP and 46P/Wirtanen, ``radar skirts" were detected, which indicates coma grains of at least 4 cm (personal communication, Ellen Howell). Specifically, for 46P/Wirtanen, we can also determine that these grains must have been moving at a minimum velocity of 15 $m s^{-1}$ away from the nucleus (personal communication, Ellen Howell). In accordance with the radar measurements, \citet{Zheltobryukhov} also found  evidence of a circumnucleus halo of dust, suggesting either diffuse outgassing, or the merge of multiple weak jets. For the large grains to be important in our measurements of radial profile slopes, there would have to be a very massive, ongoing flux of large particles. It is more likely that the smaller grains dominate the coma in visible light.

$Af\rho$, the product of albedo $A$, a filling factor $f$, and the projected radial distance $\rho$ on the plane of the sky \citep{Ahearn_1984}, is representative of dust production rates. For 45P/HMP, $Af\rho$ peaks 50 days post-perihelion at 33 cm in the $R_c$ filter (comparable to the Harris-R filter) and then decreases afterwards \citep{Moulane}. The peak in $Af\rho$ at 50 days post perihelion found by \citet{Moulane} potentially matches with a steepening of the radial profile slope, however, their time baseline does not cover as extensive a range as ours does. An increase in the dust production at the nucleus would temporarily cause the radial profile slope to be steeper than $-1$, corresponding to the small dip around 50 days post-perihelion that is visible in Figure \ref{45P_radial_slopes}.

As explained in \S \ref{grain_fragmentation_45P}, grain fragmentation can cause a steepening or a shallowing of the radial profile slope depending on the composition of the grain. Depending on the icy grain environments, such as quick fragmentation of icy grains, sublimation of icy grain, or even icy grain mantling for example, could result in different radial profile provided that the icy grains survive long enough (e.g.: \citet{Beer_2006}, \citet{Markkanen}, \citet{Davidsson_2021}). This, combined with the radar measurements, suggests that if icy grains were present in either 45P/HMP, 46P/Wirtanen, or both, an icy dust halo could be present, creating a shallowing of the radial profile slope. Comets 45P/HMP and 46P/Wirtanen have very different radial profile slopes, which suggests that the two comets have vastly different dust environments in the inner coma.

\subsection{Solar Position Angle Correlation} \label{Solar_PA_correlation_section}

\begin{figure}[ht]
\centering
\includegraphics[width=14cm, angle=270]{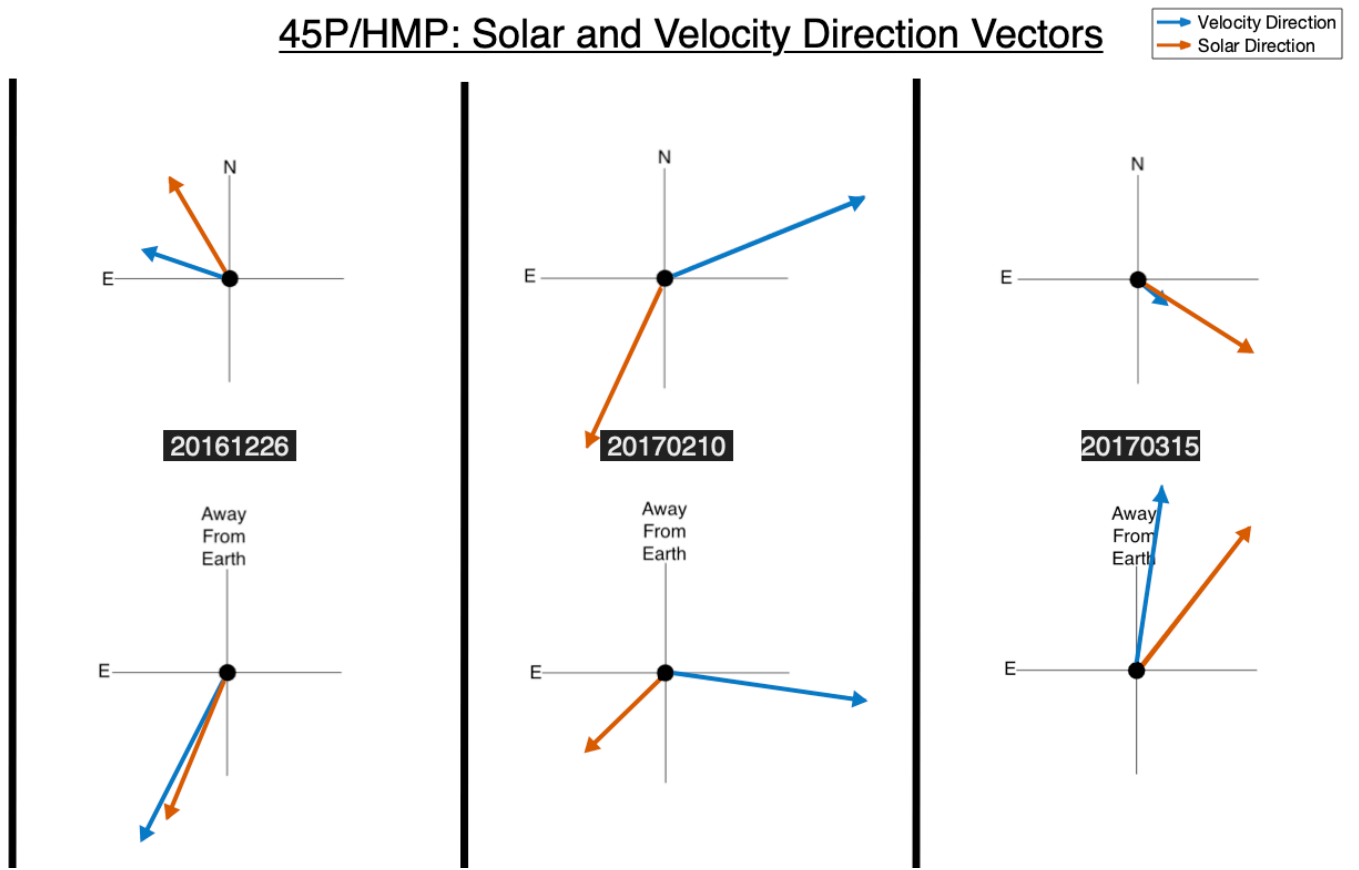}
\caption{The velocity and solar direction vectors as seen in the plane of the sky (top) and rotated 90$^{\circ}$ from the plane of the sky (bottom) for 45P/HMP on our first observation (left), our observation closest to perigee (middle), and our last observation (right). Projected North and East represent our sky plane.}
\label{45P_vectors}
\end{figure}

\begin{figure}[ht]
\centering
\includegraphics[width=14cm,angle=270]{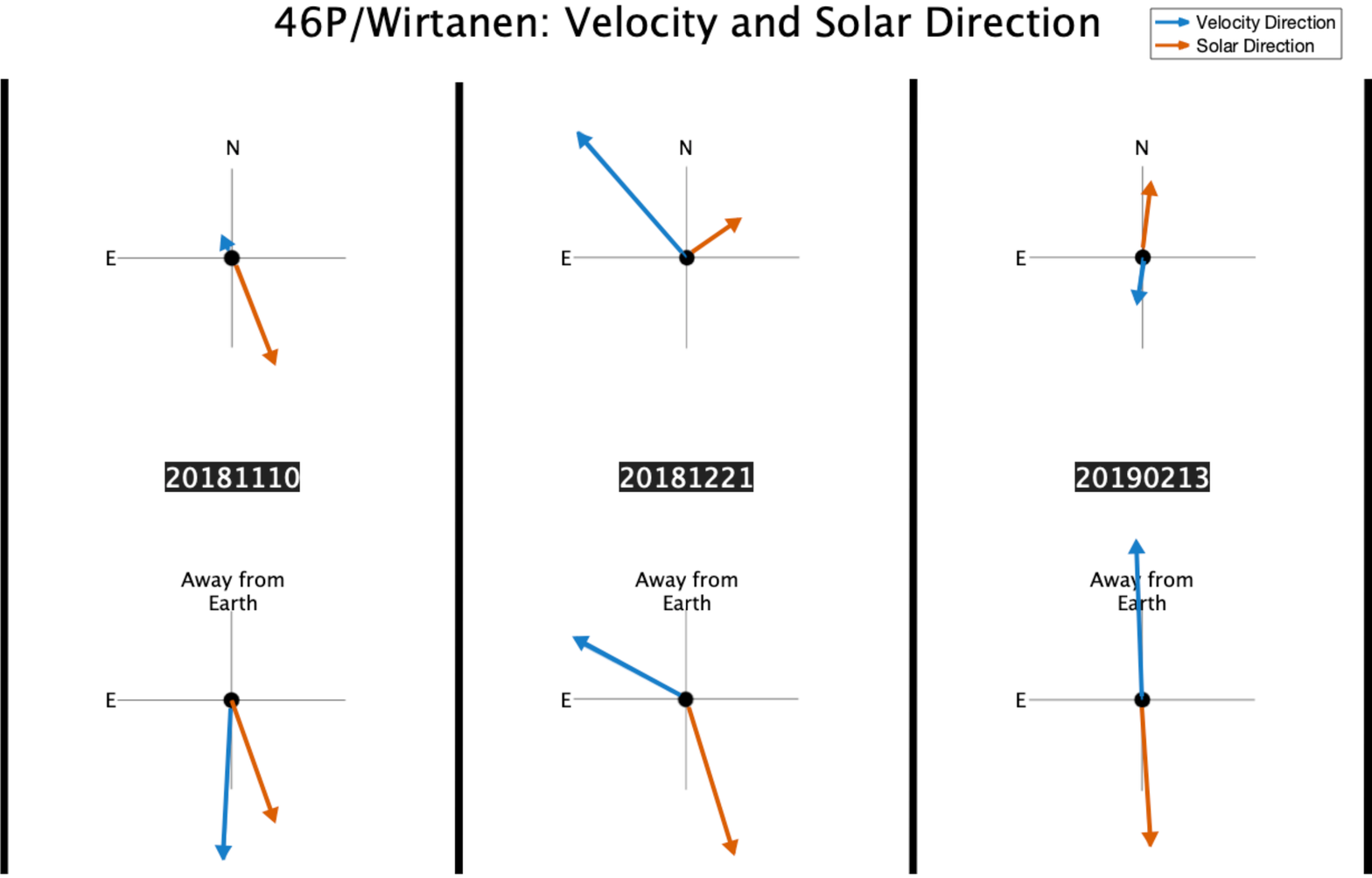}
\caption{The velocity and solar direction vectors as seen in the plane of the sky (top) and rotated 90$^{\circ}$ from the plane of the sky (bottom) for 46P/Wirtanen on our first observation (left), our observation closest to perigee (middle), and our last observation (right). Projected North and East represent our sky plane.}
\label{46P_vectors}
\end{figure}

\begin{figure}[ht]
\centering
\includegraphics[width=0.46\textwidth]{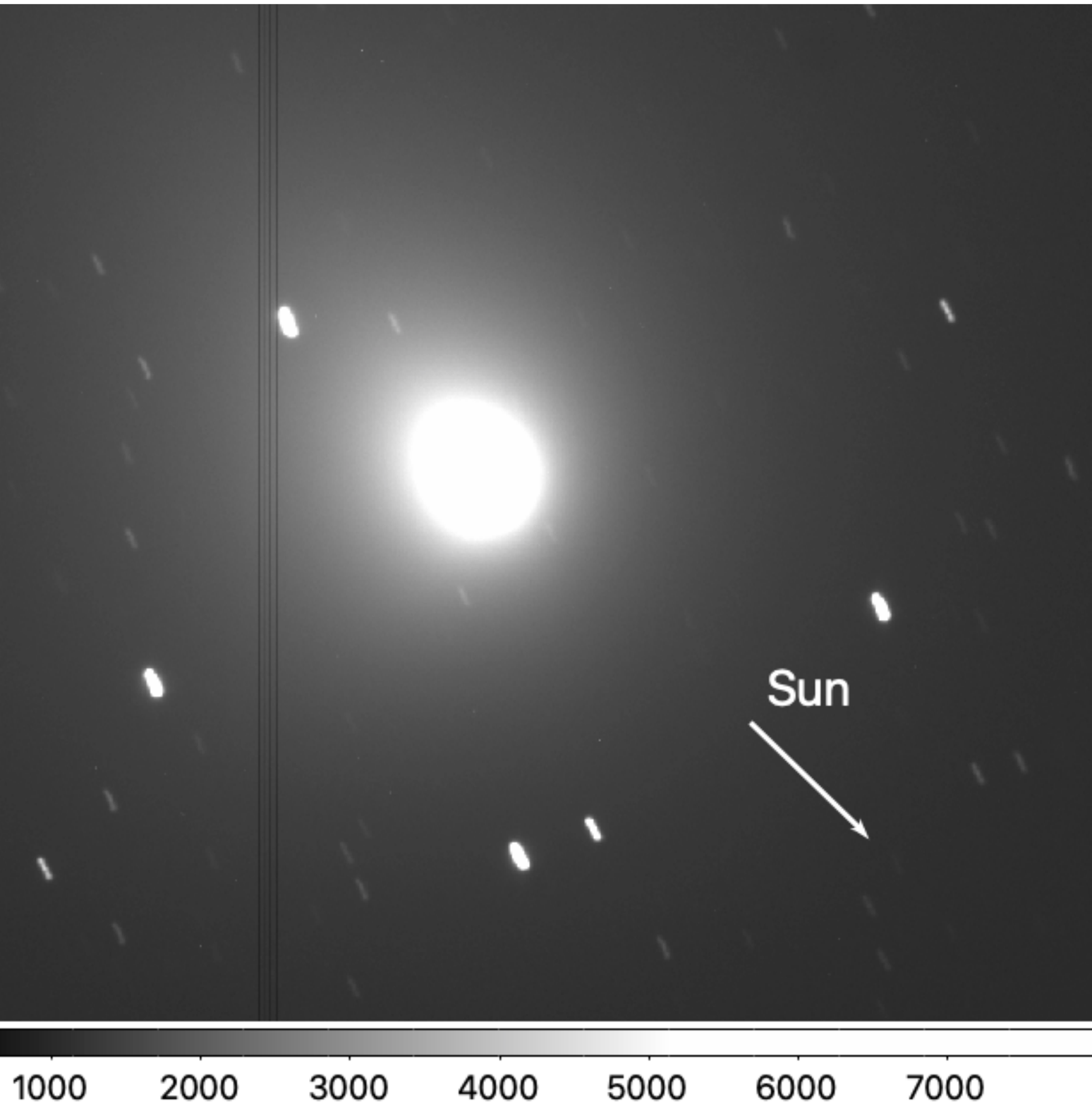}
\includegraphics[width=0.455\textwidth]{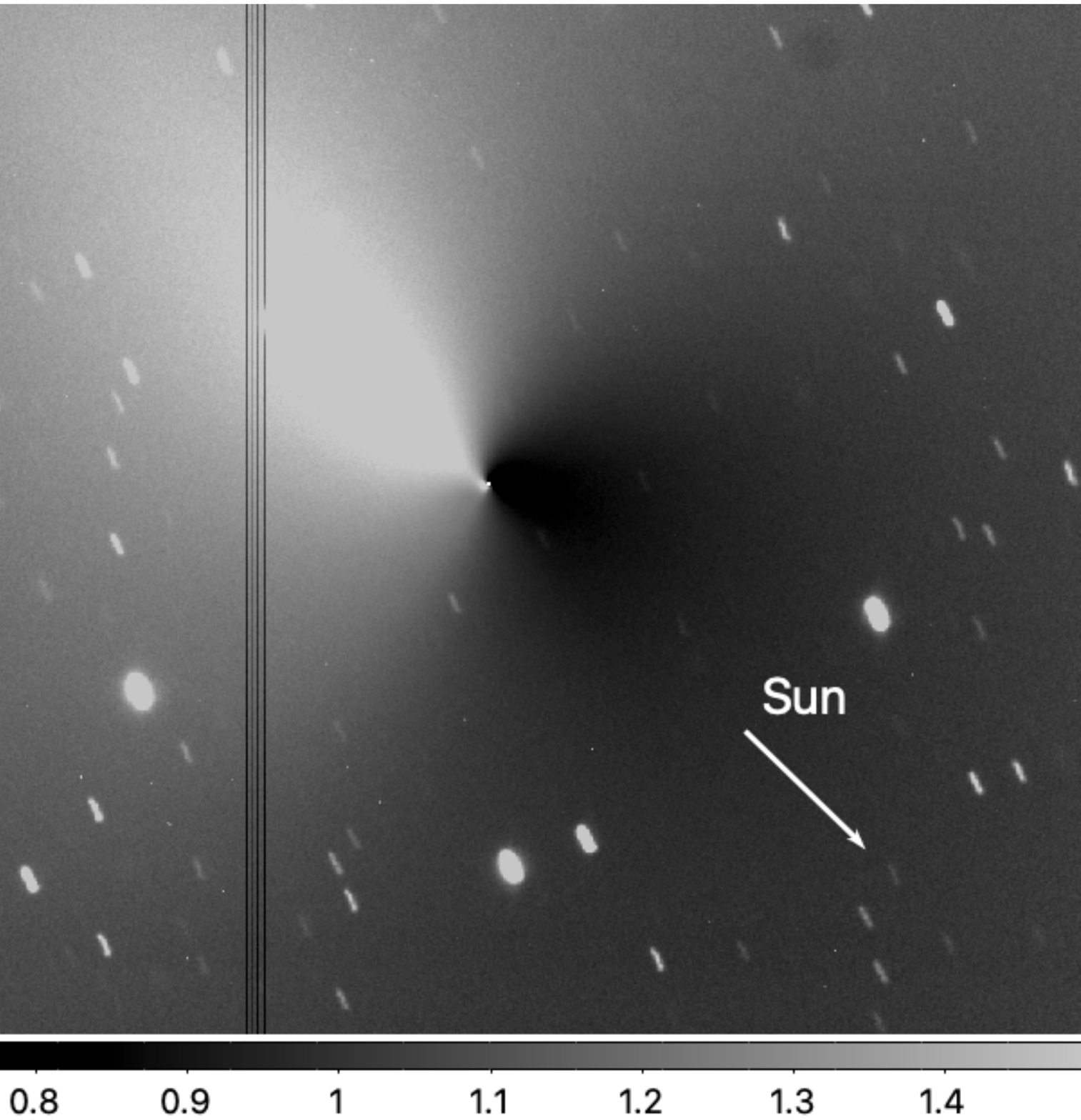}
\caption{ Left: Standard reduced image of 46P/Wirtanen on 2018-12-09. Right: Same image divided by a circularly symmetric radial profile model of the median radial profile slope measured showing a strong antisunward feature (i.e. the tail). In both images, the arrow length represents 2000 km and North is up, East is to the left. \label{46P_subtraction}}
\end{figure}

The solar position angle changes more than 180$^{\circ}$ over the course of our observations for both comets (Tables \ref{45P_dates_table} \& \ref{46P_dates_table}). Thus, it was logical to show the radial profiles with respect to the solar position angles as shown in Figures \ref{45P_az_deviation} and \ref{46P_az_deviation}. Comet 45P/HMP does not show any significant correlation with azimuthal offsets from the solar position angle. However, 46P/Wirtanen does show an overall shallower slope in the anti-sunward direction and a steeper slope in the sunward direction. This trend is visible in both the pre- and post-perihelion data. Figures \ref{45P_vectors} \& \ref{46P_vectors} show that the Sun is in the direction of the Earth for most of the 46P/Wirtanen encounter, but closer to the sky plane for 45P/HMP. This suggests that there may be more prominent changes in radial profiles due to the tail projection onto the sky plane for comet 46P/Wirtanen. We also see a weaker tail feature in Figure \ref{45P_subtraction} for 45P/HMP than in Figure \ref{46P_subtraction} for 46P/Wirtanen. A detailed grain model, outside the scope of this observational study, that includes grain fragmentation, radiation pressures, different grain compositions, etc. would be necessary to further understand the outcomes of radial profiles with respect to behaviors of grain environments. 

Essentially, in 46P/Wirtanen, we are looking at more material from the anti-sunward (tail) direction at each $\rho$, so the grains' total cross-section builds up faster potentially due to grain fragmentation, and we would observe a shallowing of the slope simply due to the projected tail feature. Additionally, because the solar phase angle is smaller, we would also observe more contamination from the direction reversal of the sunward direction moving grains towards the tailward direction due to radiation pressures.  

Contrarily, the weak anti-sunward (tail) feature of 45P/HMP is more often closer to the sky plane (see Fig. \ref{45P_subtraction}), so any potential change to the radial profile would be much smaller, and in our data, potentially be below our noise threshold. Additionally, Figure \ref{45P_subtraction} shows a weak sunward feature which does not seem to affect the radial profile slope visible in Figure \ref{45P_az_deviation}. This could simply imply that the grain in the sunward feature are behaving similarly to the grains in the rest of the coma in terms of their radial distribution even though there may simply be larger flux present.

The tailward direction radial profile slope shallowing of 46P/Wirtanen is most likely not caused by only radiation pressure in the coma. Essentially, in a spherically expanding coma subjected to radiation pressure, for the sunward direction, as you move progressively away from the nucleus, more and more grains amass because of the decrease of velocity due to radiation pressure, causing a shallowing of the slope. In the tailward direction, as you move progressively away from the nucleus, grains have had more time to accelerate due to radiation pressure, and keep accelerating, causing a steepening of the slope. This is the opposite behavior that we observe in 46P/Wirtanen. However, since we know that 46P/Wirtanen has large grains, which are less affected by radiation pressure, they have more time to disintegrate and essentially amass farther out. Depending on the mechanism of fragmentation of large grains (if they were to fragment), there could be more generations of smaller grains (i.e., micron-sized) consistently, resulting in an increase in grain cross-section, thus causing a shallower slope. 

However, it is unlikely that the tailward shallowing of 46P/Wirtanen is caused by larger grains moving along the orbit of the comet. As seen in Figure \ref{46P_vectors}, the velocity direction does not align with the solar direction (or their converse), thus the feature in the tailward direction is not correlated with the anti-velocity direction.

\section{CONCLUSION} \label{sec:conclusion}

We find that the radial profile slope of comets 45P/HMP and 46P/Wirtanen, measured over 79 and 95 days respectively, have different temporal behaviors. Comet 45P/HMP has a radial profile slope that becomes shallower with time, starting at $-1.81 \pm 0.20$ pre-perihelion and ending at $-0.35 \pm 0.16$ post-perihelion during our observations. Resolution of the inner coma, near perigee, suggest we may have been able to resolve the zone of interaction when the dust and gas were still coupled. Surprisingly, the radial profile slope near perihelion appears to be closest to -1. This suggests that there may be different dust processes occurring pre-perihelion and post-perihelion such as a change in grain composition or friability. Additionally, the transition between different processes seems gradual, suggesting that it is not caused by a single short event. On the other hand, 46P/Wirtanen's radial profile slope appears fairly constant at $-1.05 \pm 0.05$. This suggests that there is a constant behavior for the dust expansion of 46P/Wirtanen, possibly a steady state expansion of the dust coma, as described in the Fountain Model (see \S \ref{fountain_model}), and at this point, processes more complicated need not be invoked for explaining the azimuthally medianed radial profiles of 46P/Wirtanen. 

The two most interesting results from our analysis are the shallowing of 45P/HMP's slope over the apparition and the shallowing of the radial profile slope of 46P/Wirtanen in the tailward direction. Possible explanations for 45P/HMP's behavior are:
\begin{itemize}
    \item A peak in $Af\rho$ near 50 days post-perihelion and decrease right after it would cause a gradual shallowing of the radial profile slope, or
    \item A progressive change in the type of dust grains from large icy grains pre-perihelion to small non-icy grains post-perihelion.
\end{itemize}
On the other hand, the tailward shallowing of the slope for 46P/Wirtanen, that is not present in 45P/HMP, is most likely caused by a combination of:
\begin{itemize}
    \item A projection angle effect of the dust tail away from the observer's direction, and
    \item Larger grains ejected in the tailward direction are not accelerated as quickly due to radiation pressure; they may eventually fragment, cause an increase in micron-sized grains, and in turn increase the net dust grain cross-section.
\end{itemize}

Such a difference in radial profile slopes between comets 45P/HMP and 46P/Wirtanen suggests that the two have vastly different dust environments in the inner coma. This indicates that at least the upper layer of the two comets' nuclei are significantly different. Our results imply that a combination of the surface properties, total gas production, and the distribution of the source regions all play a role in the two comets' dust environments.

\begin{acknowledgments}
\section*{Acknowledgements}
We would like to thank everyone who participated in the data gathering process. This includes all students from the Lunar and Planetary Laboratory course PTYS 495B/595B (Fall 2018). We would also like to thank the additional organizers of the 4*P Campaign: Matthew Knight (United States Naval Academy) and Tony Farnham (University of Maryland). We would like the thank the Steward Observatory technical staff for the amount of time they dedicated to this project. Finally, we would like to thank the SSO grant NNX16A670G (Walt Harris) and the NESSF grant 80NSSC18K1241 (Cassandra Lejoly, PI: Walt Harris) for allowing this work to be completed. We would also like to thank the Slovak Academy of Sciences grant VEGA 2/0023/18 (Oleksandra Ivanova) and the Slovak Research and Development Agency under the Contract no. APVV-19-0072 (Oleksandra Ivanova).
\end{acknowledgments}

\end{document}